\documentclass{JHEP3}
\pdfoutput=1
\usepackage{amsmath}
\usepackage{amssymb}
\usepackage{graphicx}
\usepackage{nicefrac}
\usepackage{cite}

\DeclareFontEncoding{LGR}{}{}
\DeclareTextSymbol{\~}{LGR}{126}
\newcommand{\lyxmathsym}[1]{\ifmmode\begingroup\def\b@ld{bold}
  \text{\ifx\math@version\b@ld\bfseries\fi#1}\endgroup\else#1\fi}

\newcommand{\be}{\begin{equation}}
\newcommand{\ee}{\end{equation}}
\newcommand{\bea}{\begin{eqnarray}}
\newcommand{\eea}{\end{eqnarray}}

\makeatletter

\title{On the nature of the fourth generation neutrino and its implications}

\author{Alberto Aparici\footnote{alberto.aparici@uv.es}, 
Juan Herrero-Garc\'\i a\footnote{juan.a.herrero@uv.es},
Nuria Rius\footnote{nuria@ific.uv.es}\,
and Arcadi Santamaria\footnote{arcadi.santamaria@uv.es}  
\\
Dept.\ de F\'\i sica Te\'orica, and IFIC, Universidad de Valencia-CSIC \\ 
Edificio de Institutos de Paterna, Apt. 22085, 46071 Valencia, Spain}

\abstract{We consider the neutrino  sector of a Standard Model with 
four generations. While the three light neutrinos can obtain their masses 
from a variety of mechanisms with or without new neutral fermions, 
fourth-generation neutrinos need at least one new relatively light 
right-handed neutrino. If lepton number is not conserved this neutrino must 
have a Majorana mass term whose size depends on the underlying mechanism for 
lepton number violation. Majorana masses for the fourth-generation neutrinos 
induce relative large two-loop contributions to the light neutrino masses 
which could be even larger than the cosmological bounds. This sets strong 
limits on the mass parameters and mixings of the fourth-generation neutrinos.}  

\preprint{IFIC/12-22\\
FTUV-12-0404}

\makeatother

\begin{document}

\section{Introduction}

In the framework of the Standard Model (SM), fermions are grouped
into three families, each containing a doublet of quarks and a doublet
of leptons. The number of families is not a constructive parameter
of the theory, and it could well be four or more; for this reason,
the enlargement of the SM with new generations has been commonly considered
\cite{Holdom:2009rf}, and it has proven to help in dealing with several
problems, such as the lack of CP violation in the SM 
to explain the baryon asymmetry of the universe \cite{Hou:2008xd}
or the structure of the leptonic mass matrices \cite{SilvaMarcos:2002bz}.
The currently available SM observables, however, constrain quite tightly
the properties of such new families \cite{Nakamura:2010zzi}, and
the global electroweak fits seem to disfavour a scenario with more
than five generations \cite{He:2001tp,Novikov:2009kc}; maybe the most striking
result against the existence of additional families is the LEP measurement
of the number of neutrinos at the $Z$ peak, which
forbids more than three light neutrinos \cite{Nakamura:2010zzi}, 
but even
this can be dodged if the neutrinos of the new generations are too
heavy to be produced in $Z$ decays. All in all, the existence of
new generations is not actually excluded, and it seems worth being
considered \cite{Holdom:2009rf}, even more now that the LHC is working
and exploring the relevant mass range.

On the other hand, right-handed neutrinos constitute a common new
physics proposal, usually linked to the generation of neutrino masses.
This is particularly interesting nowadays, ever since we gathered compelling
evidence that neutrinos do have masses, that they lie well below the
other fermions' ones, and that their mixing patterns differ extraordinarily
from those of the quark sector (for a review on the matter of neutrino
masses see, for example, \cite{GonzalezGarcia:2007ib}). The most straightforward
way to construct a mass term for the neutrinos within the SM is just
to rely on the Higgs mechanism, and so to write the corresponding
Yukawa couplings; for that aim, one needs some fermionic fields which
carry no SM charge: right-handed neutrinos. However, we do not know whether 
neutrinos are Dirac or Majorana. If they are Dirac,  
the smallness of the neutrino mass scale remains 
unexplained, for it would be just a product of the smallness of the
corresponding Yukawa couplings. In order to provide such an explanation,
many models and mechanisms have been proposed: in the so-called see-saw
models, the lightness of the neutrino mass scale is a consequence of the
heaviness of another scale. 
For instance, this scale is the  
lepton-number-violating (LNV) Majorana mass of the extra 
right-handed neutrinos in type I see-saw
\cite{Minkowski:1977sc,GellMann:1980vs,Yanagida:1979as,Mohapatra:1979ia}. On the other hand, 
radiative models propose that neutrino  masses are originated via suppressed,
high-order processes~\cite{Zee:1980ai,Zee:1985id,Babu:1988ki,Ma:2006km}. 
Although some of these proposals do not require right-handed neutrinos, for the sake of 
generality it is
a good idea to consider their possible involvement in the generation of neutrino masses.

In this work we aim to discuss the naturality of the various scenarios
arising when new generations and right-handed neutrinos are brought
together. Several previous works have considered such association,
either explicitly, in order to provide a mechanism for mass generation,
or implicitly, when assuming Dirac neutrinos in their 
analyses~\cite{Babu:1988ig,Babu:1988wk,Hill:1989vn,Mohapatra:1993vm,Carpenter:2010dt,Rajaraman:2010ua,Lenz:2010ha,Aparici:2011nu,Lenz:2011gd,Schmidt:2011jp}. 
We argue that unless a symmetry is invoked which separates the new family
from the first three, the coexistence of both Dirac and Majorana neutrinos
is not stable under radiative corrections and doesn't seem natural~\cite{Aparici:2011eb,Schmidt:2011jp}.
Furthermore, the presence of a fourth family plus a right-handed Majorana neutrino triggers the
generation of Majorana masses for the light species through a well-known
mechanism \cite{Petcov:1984nz,Branco:1988ex,Grimus:1989pu,Babu:1988ig,Babu:1988wk,Aparici:2011nu,Schmidt:2011jp}; the upper 
bounds on the
light neutrino masses can thus be translated into bounds on the mixings
with the new, heavy generations. 

This paper is structured as follows.
In section \ref{SM3} we start by reviewing the different mechanisms which can provide light neutrino masses. In section \ref{SM4} we discuss the naturalness of those mechanisms 
to generate the fourth-family neutrino mass and conclude that at least one right-handed 
neutrino is needed. Assuming that light neutrinos are Majorana, we  
use naturalness arguments to provide a lower bound on the Majorana mass of the 
right-handed neutrino.
In section \ref{4gto2gmasses} we consider a minimal four generation SM with only one 
relatively light right-handed neutrino and Majorana masses for light neutrinos parametrized 
by the Weinberg operator~\cite{Weinberg:1979sa,Weldon:1980gi}. We describe the radiative, two-loop contribution
of the heavy fourth-family neutrinos to the light neutrino mass matrix. 
In section \ref{pheno} we discuss
the phenomenological consequences  of this
minimal four-generation scenario  with heavy Majorana neutrinos
(lepton flavour violation,  universality bounds, light neutrino masses, 
neutrinoless double beta decay,\ldots),
and we conclude in section \ref{conclusions}.
Appendix A is devoted to describe an explicit example in which a finite 
Majorana mass for the fourth-generation right-handed neutrino is radiatively generated. 

\section{Light neutrino masses} \label{SM3}

The huge hierarchy between neutrino masses and those of all other 
fermions has triggered the appearance of 
many different mechanisms to explain the lightness of neutrinos. 
Here we briefly review some of these mechanisms, 
with special emphasis on the frameworks that are able to explain 
neutrino masses including a fourth generation, which
will be discussed in the next section.

\subsection{Dirac masses}
If there are right-handed neutrinos and a conserved global symmetry
(for instance $B-L$) prevents them from having a Majorana mass, neutrinos are
Dirac particles, as all other fermions in the SM. However, in this scenario
there is no explanation for the smallness of neutrino masses, having to
impose by hand extremely tiny Yukawa couplings, approximately 6 (11)
orders of magnitude smaller than the electron (top) one.
Therefore, although in principle it is possible, 
a Dirac nature does not seem the most natural option for neutrinos
(but see, for example, \cite{SilvaMarcos:2002bz}, for a proposal
in this direction which avoids tiny Yukawas).

\subsection{Seesaw}

Seesaw models are minimal extensions of the SM which can naturally lead 
to tiny (Majorana) neutrino masses, keeping  the SM 
gauge symmetry, $SU(3)_\mathrm{C} \otimes SU(2)_\mathrm{L}
\otimes U(1)_\mathrm{Y}$ and 
renormalizability, but giving up the (accidental) lepton number conservation
of the SM. 
Let's explain briefly the different types to fix notation.

\subsubsection{Type I:  fermionic singlets}

In type I see-saw~\cite{Minkowski:1977sc,Ramond:1979py,GellMann:1980vs,Yanagida:1979as,Mohapatra:1979ia}, 
 $n$ SM fermionic singlets with zero hypercharge
are added to the SM; these have the quantum numbers of right-handed 
neutrinos, and can be denoted by $\nu_{\mathrm{R} i}$. Note that
to explain neutrino data, which requires al least two massive neutrinos,
a minimum of two extra singlets are needed.
Having no charges under the SM, Majorana masses
for right-handed neutrinos are allowed by gauge invariance, so the 
new terms in the Lagrangian are:

\be
\mathcal{L}_{\nu_\mathrm{R}} = i \, \overline{\nu_\mathrm{R}} \gamma^{\mu} 
	\partial_{\mu} \nu_\mathrm{R} - \left( \dfrac{1}{2} 
	\overline{\nu_\mathrm{R}^\mathrm{c}} M \nu_\mathrm{R} +
	\overline{\ell} \, \tilde{\phi} \, Y \,\nu_\mathrm{R} +
	\mathrm{H. c.} \right)
\, ,
\ee
where $\ell$ and $\phi$ are respectively the lepton and Higgs SM  doublets,
$\tilde{\phi} = i \, \tau_{2} \, \phi^{*}$ with $\tau_2$, the second 
Pauli matrix, acting on the $SU (2)_\mathrm{L}$ indices,
$M$ is a $n \times n$ symmetric matrix, $Y$ is a general $3 \times n$ 
matrix and we have omitted flavour indices for simplicity.
After spontaneous symmetry breaking (SSB), 
$\langle \phi \rangle = v_\phi$ with $v_\phi =$ 174 GeV, 
the neutrino mass terms are 
given by
\be
\mathcal{L}_{\nu\: \mathrm{mass}} = -\dfrac{1}{2} \, 
	\begin{pmatrix} 
		\overline{\nu_\mathrm{L}} & \overline{\nu_\mathrm{R}^\mathrm{c}}
	\end{pmatrix} \, 
	\begin{pmatrix}
		0 & m_\mathrm{D} \\
		m_\mathrm{D}^\mathrm{T} & M
	\end{pmatrix} \,
	\begin{pmatrix}
		\nu_\mathrm{L}^\mathrm{c} \\
		\nu_\mathrm{R}
	\end{pmatrix} + \mathrm{H.c.} \ , 
\ee
where $m_D = Y v_{\phi}$.
The mass scale for right-handed neutrinos is in principle free, however if 
$M \gg m_\mathrm{D}$, upon block-diagonalization one obtains 
$n$ heavy leptons which are mainly SM singlets, with masses 
$\sim M$, and the well-known 
see-saw formula for the effective light neutrino Majorana mass matrix, 
\be
m_{\nu} \simeq -m_\mathrm{D} \, M^{-1} \, m_\mathrm{D}^\mathrm{T} \, ,
\ee
which naturally explains the smallness of light neutrino masses as a 
consequence of the presence of heavy SM singlet leptons.

\subsubsection{Type II: scalar triplet}\label{sec:scalar-triplet}

The type II see-saw~\cite{Konetschny:1977bn,Cheng:1980qt,Lazarides:1980nt,Magg:1980ut,Schechter:1980gr} only adds to 
the SM field content one scalar triplet 
with hypercharge $Y=1$ (we adopt the convention that $Q = Y + T_3$) and 
assigns to it lepton number $L=-2$. 
In the doublet representation of $SU(2)_\mathrm{L}$ the triplet can
be written as a $2 \times 2$ matrix, whose components are
\be
\chi = \begin{pmatrix}
			\chi^{+}/\sqrt{2} & \chi^{++} \\
			\chi_{0} & - \chi^{+}/\sqrt{2} 
		\end{pmatrix} \, .
\ee
Gauge invariance allows a Yukawa coupling of the scalar triplet to 
two lepton doublets, 
\be \label{Lst}
\mathcal{L}_{\chi} =  \left( (Y^\dagger_\chi)_{\alpha \beta} \, \overline{\tilde \ell}_\alpha 
		\chi \ell_\beta + \mathrm{H. c.} \right) - V(\phi,\chi) \, ,
\ee
where $Y_\chi$ is a symmetric matrix in flavour space, and $\tilde \ell = i
\tau_2 \, \ell^\mathrm{c}$. 
The scalar potential has, among others, the following terms:
\bea
V (\phi, \chi) = m_\chi^2 \, \mathrm{Tr}[ \chi \chi^\dagger] -
			\left( \mu \,\tilde \phi^{\dagger} \chi^{\dagger} \phi 
			+ \mathrm{H.c.} \right) + \ldots
\label{eq:triplet-potential}
\eea
The $\mu$ coupling violates lepton number explicitly, and it induces a 
vacuum expectation value (VEV) for the triplet via the VEV of the doublet,
even if $m_\chi > 0$.
In the limit $m_\chi \gg v_\phi$ this VEV can be approximated by:
\be
\left< \chi \right> \equiv v_\chi 
		\simeq \cfrac{\mu  v_\phi^{2}}{m_{\chi}^{2}} \, ;
\ee
then, the Yukawa couplings in equation \eqref{Lst} lead to a Majorana mass
matrix for the left-handed neutrinos
\begin{equation}
m_\nu = 2 Y_\chi v_\chi = 2 Y_\chi  \cfrac{\mu v_\phi^{2}}{m_{\chi}^{2}} \, .
\label{eq:triplet-numass}
\end{equation}
Neutrino masses are thus proportional to both $Y_\chi$ and $\mu$. 
Such dependence can be understood from the Lagrangian, since the breaking 
of lepton number $L$ results from the simultaneous presence of the Yukawa 
and $\mu$ couplings. As long as $m_{\chi}^{2}$ is positive and large, 
$v_\chi$ will be small, in agreement with
the constraints from the $\rho$ parameter, $v_\chi \lesssim 6 \; \mathrm{GeV}$
\cite{Kanemura:2012rs}.\footnote{
This bound is calculated after the inclusion of the one-loop corrections to the 
$\rho$ parameter,
and is slightly looser than other previously obtained from electroweak
global fits (see, for example, \cite{delAguila:2008ks}). Note also that 
the authors of \cite{Kanemura:2012rs} use 
a different normalisation for the VEV, and hence the difference between
their value and the one we present here.
}
Moreover, the parameter $\mu$, which has dimensions of mass, 
can be naturally small, because in its absence lepton number is recovered, 
increasing the symmetry of the model.

\subsubsection{Type III: fermionic triplets}

In the type III see-saw model~\cite{Foot:1988aq,Ma:2002pf}, the SM is extended by fermion $SU(2)_{\mathrm{L}}$ 
triplets $\Sigma_{\alpha}$ with zero hypercharge. As in type I, at least two 
fermion triplets are needed to have two non-vanishing light neutrino masses.
We choose the spinors $\Sigma_\alpha$ to be right-handed under Lorentz transformations and
write them in $SU(2)$ Cartesian components $\vec{\Sigma}_{\alpha}=(\Sigma_\alpha^1,\Sigma_\alpha^2,\Sigma_\alpha^3)$. 
The Cartesian components can be written in terms of charge eigenstates as usual 
\begin{equation}
\Sigma^+_\alpha=\frac{1}{\sqrt{2}} (\Sigma^1_\alpha-i\Sigma^2_\alpha)\ ,\qquad 
\Sigma^0_\alpha=\Sigma^3_\alpha\ , \qquad
\Sigma^-_\alpha=\frac{1}{\sqrt{2}} (\Sigma^1_\alpha+i\Sigma^2_\alpha)\ ,
\end{equation}
and the charged components can be further combined into negatively charged Dirac fermions 
$E_\alpha=\Sigma_\alpha^-+\Sigma_\alpha^{+\mathrm{c}}$.  
Using standard four-component notation the new terms in the Lagrangian are given by
\begin{equation}
\mathcal{L}_{\Sigma} = i \,\overline{\vec{\Sigma}_\alpha} \gamma^{\mu}
		D_{\mu} \cdot\vec{\Sigma}_\alpha - 
		\left( \cfrac{1}{2} \, M_{\alpha\beta}\overline{\vec{\Sigma}^{\mathrm{c}}_\alpha}\cdot \vec{\Sigma}_\beta +
		Y_{\alpha\beta} \, \overline{\ell_\alpha} \, \left(\vec{\tau}\cdot\vec{\Sigma}_\beta\right) 
		\tilde \phi + \mathrm{H.c.} \right) \, ,
\end{equation}
where $Y$ is the Yukawa coupling of the fermion triplets to the SM lepton 
doublets and the Higgs, and $M$ their Majorana mass matrix, which  
can be chosen to be diagonal and real in flavour space.

After SSB the neutrino mass matrix can be written as
\be
\mathcal{L}_{\nu\: \mathrm{mass}} = - \cfrac{1}{2} \,
		\begin{pmatrix}
			\overline{\nu_\mathrm{L}} & \overline{\Sigma^{0\mathrm{c}}}
		\end{pmatrix} \,
		\begin{pmatrix}
			0 & m_{\mathrm{D}} \\
			m_{\mathrm{D}}^{\mathrm{T}} & M
		\end{pmatrix} \,
		\begin{pmatrix}
			\nu_{\mathrm{L}}^{\mathrm{c}} \\
			\Sigma^{0}
		\end{pmatrix}
		+ \mathrm{H. c.} \, , 
\ee
which is the same as in the type I see-saw just replacing the singlet 
right-handed neutrinos by the neutral component of the triplets, 
$\Sigma^0_{\alpha}$, and therefore leads to a light neutrino Majorana mass matrix 
\be
m_{\nu} \simeq -m_{\mathrm{D}}\, M^{-1} \, m_{\mathrm{D}}^{\mathrm{T}}\, .
\ee
However, since the triplet has also charged components with the same 
Majorana mass, in this case there are stringent lower bounds on the new 
mass scale, $M \gtrsim 100 \; \mathrm{GeV}$.

\subsection{Others}

Here we briefly summarize non minimal mechanisms which also  lead to 
Majorana light neutrino masses. Most of these models do not include right-handed neutrinos
and are designed to obtain tiny Majorana masses for the left-handed
SM neutrinos, so we can anticipate that they will not be appropriate for the fourth 
generation.

\begin{description}
\item[a)] \textbf{Radiative mechanisms.}
Small Majorana neutrino masses may also be induced by radiative 
corrections~\cite{Zee:1980ai,Zee:1985id,Babu:1988ki,Ma:2006km}. Typically, 
on top of loop  factors of at least $1/(4\pi)^2$, there are additional 
suppressions due to couplings or ratios of masses, leading to the observed light neutrino 
masses with a new physics scale not far above the electroweak one.

\item[b)] \textbf{Supersymmetry.}
There is an intrinsically supersymmetric way of breaking lepton number 
by breaking the so-called R parity~\cite{Aulakh:1982yn,Hall:1983id,Lee:1984kr,Lee:1984tn,Ellis:1984gi,Ross:1984yg,Dawson:1985vr,Santamaria:1987uq,Masiero:1990uj,Hirsch:2000ef}
(for a review see \cite{Barbier:2004ez}). 
In this scenario,  the SM doublet neutrinos 
mix with the neutralinos, i.e., the supersymmetric (fermionic) partners of 
the neutral gauge and Higgs bosons. As a consequence, Majorana masses for 
neutrinos (generated at tree level and at one loop)
are naturally small 
because   they are proportional to the small R-parity-breaking parameters.

\end{description}

\subsection{Weinberg operator}\label{sec:weinbeg-operator} 

As we have mentioned, it does not seem very natural that neutrinos are 
Dirac particles; assuming that they are Majorana, we will be often 
interested in abstracting from the actual mechanism of mass generation.
In such case, if the light degrees of freedom are those of the SM 
we can parametrise the Majorana masses in terms of the well-known
dimension 5 Weinberg operator
\footnote {In supersymmetric models, $\tilde\phi = H_u$, since there are two Higgs doublets.} \cite{Weinberg:1979sa,Weldon:1980gi}:
\begin{equation}
\mathcal{L}_{5}= \cfrac{1}{2} \, \frac{c_{\alpha\beta}}{\Lambda_W}
		(\overline{\ell_\alpha}\tilde\phi) \,(\phi^{\dagger}\tilde{\ell}_\beta)
		+ \mathrm{H. c.}\, ,
\label{eq:weinberg-operator}
\end{equation}
where
$\Lambda_W \gg v_\phi$ is the scale of new physics and 
$c_{\alpha\beta}$ are model-dependent coefficients with flavour structure, which 
in some models can carry additional suppression due to
loop factors (as is the case in radiative mechanisms) and/or 
ratios of mass parameters (for instance in type II see-saw $c \propto \mu/m_{\chi}$). In those cases
we will assume that $\Lambda_W$ is directly related to the masses of the new particles and absorb all 
suppression factors in $c_{\alpha\beta}$. 
 
Upon electroweak symmetry breaking, the Weinberg operator leads 
to a Majorana mass matrix for the light neutrinos of the form 
\begin{equation}
m_{\nu} = c \, \frac{v_\phi^{2}}{\Lambda_W}\, .
\label{eq:weinberg-numass}
\end{equation}
Notice that if $c_{\alpha\beta}$ is suppressed, the scale $\Lambda_W$ does not
need to be extremely large in order to fit light neutrino masses and, thus, the Weinberg
operator can parametrize a variety of Majorana neutrino mass models, including those
with masses generated radiatively.  

\section{Fourth-generation neutrino masses} \label{SM4}

If there exists a fourth generation, the fourth-generation neutrinos
must be massive (with masses $\gtrsim m_{Z}/2$ in order to avoid the
strong limits for the number of active neutrinos found at LEP). In
principle all mass mechanisms available for the light neutrinos are
also available to the fourth-generation neutrinos, however, the fact
that they must be quite massive changes completely the discussion of
the naturalness of the different mechanisms. Let us discuss them:

\begin{description}
\item [{a)}] \textbf{Dirac masses.}

Since the fourth generation must be at the
electroweak scale, this mechanism of mass generation is quite natural
for the fourth-generation neutrinos as long as lepton number is conserved.

\item [{b)}] \textbf{Fermionic singlets with Majorana mass.}

If lepton number is not conserved
there is no reason to forbid a Majorana mass term for right-handed neutrinos
(see the discussion below). However if $m_{R}\gg m_{D}$ the see-saw formula
applies and the spectrum contains a relatively light, almost-active neutrino
with mass $m_{4}\sim m_{D}^{2}/m_{R}$, which must be heavier than $m_{Z}/2$.
Therefore $m_{R}<m_{D}^{2}/m_{Z}$ and the mass of the right-handed
neutrino cannot be much larger than the electroweak scale. 
On the other hand, if $m_{R} \ll m_{D}$
there are two almost degenerate neutrinos and we are in the pseudo-Dirac
limit, which does not pose any problem.

\item [{c)}] \textbf{Scalar triplet.}

In principle, as in the case of light neutrinos, scalar triplets could also be used 
to obtain Majorana masses for the fourth-generation neutrinos.
However, the  strong
limits on the triplet's VEV coming from the $\rho$ parameter
 $v_\chi \lesssim 6 \; \mathrm{GeV}$
will yield fourth-generation neutrino masses too small. This
limit could be relaxed a bit if radiative corrections to the $\rho$
parameter coming from triplet masses are large and such that cancel
in part the deviations induced by the triplet's VEV, but this will require quite
a high degree of fine tuning among rather different quantities. Therefore,
this mechanism alone is not a natural mechanism for the fourth-generation neutrino masses.

\item [{d)}] \textbf{Fermionic triplets.}

This is similar to b), but 
together with the right-handed neutrinos there come new charged fermions
degenerate with the neutral component. Since production limits tell
us that the charged fermions must be heavier than about $100 \; \mathrm{GeV}$,
in this case the pseudo-Dirac limit is not possible. Moreover these
new fermions cannot be extremely heavy, because otherwise the active
neutrino will be too light. We conclude that this mechanism is viable
but much more constrained than b). 

\item [{e)}] \textbf{Radiative mechanisms and SUSY with broken R parity.}

Neutrino masses in these models are strongly suppressed 
with respect to the electroweak scale by  
either loop factors, couplings and/or ratios of masses. 
Therefore  they are not viable for the fourth generation.  

\item [{f)}] \textbf{Weinberg operator.}

In principle the Weinberg operator could
also be used to give Majorana masses to the fourth-generation neutrinos.
However, it will provide masses $\mathcal{O}( \nicefrac{v_\phi^2}{\Lambda_W} )$
which should be $\gtrsim m_Z/2$, so the scale of new physics
$\Lambda_W$ can not be much larger than the electroweak scale $v_\phi$ 
and the effective theory does not make sense. Therefore
the Weinberg operator does not provide a useful parametrization of
the fourth-generation neutrino mass.

\end{description}

We can therefore conclude that only a), b) (which includes a) in some limit) and 
possibly d) are good mechanisms for the fourth-generation neutrino masses. 
It seems then  that to describe correctly the fourth-generation
neutrino one needs at least one right-handed neutrino 
(either SM singlet or triplet) which has
standard Dirac couplings to the doublets. 
If this RH neutrino is a SM triplet, we have seen that its Majorana mass is 
in the range 100 GeV $\lesssim m_R \lesssim$ few TeV.

However, if the  right-handed neutrino is a SM singlet  it
could have a very small or even vanishing Majorana mass term. 
Is it natural to have Dirac neutrinos for the fourth generation? The
answer is simple: yes, provided there is a symmetry that protects them from
acquiring a Majorana mass term. 
This is not the situation if the light neutrinos are Majorana, 
as most of the SM extensions that we considered in section \ref{SM3}, 
and they can mix freely with the heavy fourth
family. 
We argue that in such a case a Majorana mass term  for the fourth right-handed neutrino
should be allowed just on symmetry grounds, and in fact, 
based on naturality arguments, a lower bound for this Majorana mass can be given.

\FIGURE{
	\includegraphics[width=0.6\textwidth]{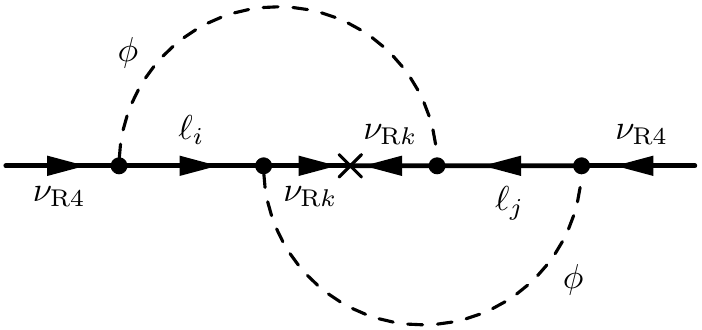}
	\caption{The two-loop process that provides a Majorana mass for the 
		fourth-generation right-handed neutrino in the framework of 
		type I see-saw. The indices $i$ and $j$ represent any of the
		four families; the index $k$, however, represents only the
		right-handed neutrinos associated to the generation of masses
		for the light families, $k = 1, 2, 3$.
		} \label{2loop-seesaw1}
}

Let us consider first the case in which the three light neutrinos obtain 
their masses via a type-I see-saw containing heavy right-handed neutrinos
with masses $m_{\mathrm{R}k}$ (with $k=1,2,3$) of the order of $10^{12}$--$10^{15}$~GeV. 
Since lepton
number is not conserved it is natural to consider a Majorana mass term for 
the fourth right-handed neutrino, $\nu_{\mathrm{R} 4}$. However, in order to satisfy 
the LEP bounds on the number of light active neutrinos, $m_{\mathrm{R}4}$ should be,
at most, of the order of a few TeV and, therefore, much smaller than $m_{\mathrm{R}k}$.
Thus, one might think that perhaps it is more natural to set directly 
$m_{\mathrm{R}4}=0$ and consider only Dirac neutrinos for the fourth generation. 
The question that arises then is whether this choice is stable or not under radiative 
corrections and what is the natural size one might expect for $m_{\mathrm{R}4}$, since
setting $m_{\mathrm{R}4}=0$ does not increase the symmetries of the Lagrangian.
The answer can be obtained from the diagram in figure~\ref{2loop-seesaw1} which gives
a logarithmically divergent contribution to $m_{\mathrm{R}4}$ induced by the presence
of the three heavy Majorana neutrino masses, $m_{\mathrm{R}k}$
\cite{Aparici:2011eb,Schmidt:2011jp}. Thus, above the 
$m_{\mathrm{R}k}$ scale, $m_{\mathrm{R}4}$ and $m_{\mathrm{R}k}$  mix
under renormalization and do not run independently. Therefore, even if one finds
a model in which $m_{\mathrm{R}4}=0$ at some scale $\Lambda_C > m_{\mathrm{R}k}$,
$m_{\mathrm{R}4}$ will be generated by running from $\Lambda_C$ to $m_{\mathrm{R}k}$.
This running can easily be estimated from the diagram in figure~\ref{2loop-seesaw1} and,
barring accidental cancellations, one should require
\be
m_{\mathrm{R} 4} \gtrsim \frac{1}{(4\pi)^{4}} \sum_{i j k } 
	Y_{i4} Y_{i k}^* m_{\mathrm{R} k} Y_{j k}^* Y_{j4}\ln(\Lambda_C/m_{\mathrm{R} k})
\gtrsim \frac{1}{(4\pi)^{4}} \sum_{i j k } 
	Y_{i4} Y_{i k}^* m_{\mathrm{R} k} Y_{j k}^* Y_{j4}\, ,
\label{eq:mr-seesaw1}
\ee
where $i,j=1,2,3,4$, $k=1,2,3$ and in the last step we have taken 
$\ln(\Lambda_C/m_{\mathrm{R} k}) \gtrsim 1$. 
Of course, given a particular renormalizable model yielding $m_{\mathrm{R}4}=0$ at tree level 
(see appendix~\ref{app-A} for an explicit example) one should be able to compute the full 
two-loop mass  $m_{\mathrm{R}4}$, which will be finite and will contain the logarithmic 
contributions we have just discussed.

Eq.~\eqref{eq:mr-seesaw1} sets the lower bound that we had announced. Let us now
estimate its value;
bearing in mind that in type I see-saw the light neutrino masses are given 
by\footnote{Notice that by integrating out the three heavy right-handed neutrinos 
we obtain a Majorana neutrino mass matrix for the four active neutrinos which
is of the order of the light neutrino masses.} 
$(m_{\nu})_{ij} \sim \sum_k Y_{ik} Y_{jk} v_{\phi}^{2}/m_{\mathrm{R} k}$. Then,
by taking all $m_{\mathrm{R}k}$ of the same order we can rewrite the bound as
\begin{equation} \label{eq:mr-seesaw}
		m_{\mathrm{R} 4} 
  \gtrsim \sum_{ij}\frac{Y_{i4} (m^*_\nu)_{ij} Y_{j4}}{(4\pi)^{4}}
  \frac{m^2_{\mathrm{R} k}}{v_{\phi}^{2}}\, .
\end{equation}
To give a conservative estimate we consider only the contribution of the first three
generations because we expect their Yukawa couplings to the fourth right-handed neutrino 
to be somewhat suppressed due to universality and LFV constraints \cite{Babu:1988ig,Babu:1988wk,Buras:2010cp,Aparici:2011nu} 
(say, $Y_{k 4} \sim 10^{-2}$).
Once we fix the neutrino masses and the Yukawa couplings between the fourth-generation 
neutrino and the first three, $m_{\mathrm{R} 4}$ grows 
quadratically with $m_{\mathrm{R} k}$. 
For $m_\nu=0.01 \; \mathrm{eV}$ and $Y_{k 4} = 0.01$ we obtain that 
$m_{\mathrm{R} 4}$ is of order keV, GeV, PeV for 
$m_{\mathrm{R} k} = 10^9, 10^{12}, 10^{15} \; \mathrm{GeV}$, respectively.
The contribution of the fourth active neutrino is not necessarily suppressed by the 
Yuwawa couplings and, in principle, by using it, even more restrictive bounds on $m_{\mathrm{R} 4}$  could be set. 
However, as $(m_\nu)_{44}$ is model-dependent\footnote{In this case, $(m_\nu)_{44}$ is the see-saw mass induced 
by only the three heavy right-handed neutrinos, thus, 
it could even be zero if the Yukawas between the fourth lepton doublet and the three right-handed neutrinos vanish 
for some reason.}, we keep the most conservative bound.

\FIGURE{
	\centering
	\includegraphics[width=0.6\textwidth]{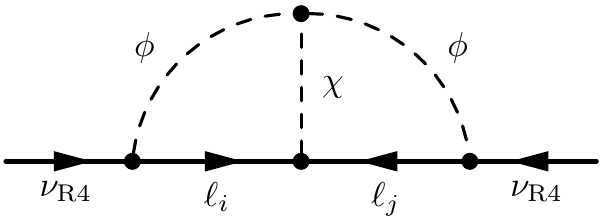}
	\caption{The process that provides a Majorana mass for the right-handed
		neutrino associated to the fourth generation in the framework of
		type II see-saw.} \label{2loop-seesaw2}
}

Let us consider now the case in which the three light neutrinos obtain their masses
through the type II see-saw mechanism (see section~\ref{sec:scalar-triplet}), {\em i.e.},
through their coupling to a scalar triplet, $\chi$, which develops a VEV. 
As discussed in section~\ref{SM4}, this triplet cannot be the only source Majorana masses
for the fourth-generation neutrinos and, at least, one right-handed neutrino is needed.
We will assume then that there is a right-handed neutrino
which has Yukawa couplings to the four SM doublets. In this scenario one can easily see that 
the right-handed neutrino
will acquire, at two loops (as seen in figure \ref{2loop-seesaw2}), 
a Majorana mass. This just reflects the fact that the right-handed neutrino mass 
$m_{\mathrm{R}4}$ and the trilinear coupling of the triplet, $\mu$, mix under 
renormalization. Applying the same arguments used in the case of see-saw type I for 
light neutrino masses and the  estimate of the diagram in figure \ref{2loop-seesaw2} 
we can write
\be
m_{\mathrm{R}4} \gtrsim \frac{\mu}{(4\pi)^{4}} 
	\sum_{ij} Y_{i4} (Y^*_\chi)_{ij} Y_{j4}\, ,
\ee
where $Y_\chi$ are the Yukawa couplings of the triplet to the lepton doublets
and, as before, we have taken $\ln(\Lambda_C/m_{\chi}) \gtrsim 1$.
As in the type I see-saw case the result can also be expressed in terms of the light neutrino masses 
$(m_{\nu})_{ij} \sim (Y_\chi)_{ij} \mu v_{\phi}^{2}/m_{\chi}^{2}$; thus
\begin{equation} \label{eq:mr-triplet}
m_{\mathrm{R}4} 
 \gtrsim \sum_{ij}\frac{Y_{i4} (m^*_\nu)_{ij} Y_{j4}}{(4\pi)^{4}}
  \frac{m^2_\chi}{v^2_{\phi}} \, , 
\end{equation}
which shows a similar structure to that obtained for type I see-saw,
eq. \eqref{eq:mr-seesaw}. The same result is obtained for type III see-saw,
whose couplings are analogous to those of type I.

The similarity of the two results  suggests that bounds
of this type are quite general and should appear in all kinds of
four-generation models with light Majorana neutrinos. In fact, as discussed in
section~\ref{sec:weinbeg-operator},
light Majorana neutrino masses 
can be parametrized in many models by means of the 
Weinberg operator, eq.~(\ref{eq:weinberg-operator}), which yields neutrino masses
given by  eq.~(\ref{eq:weinberg-numass}). Then, one could draw a two-loop diagram analogous 
to the diagrams in figures
\ref{2loop-seesaw1} and \ref{2loop-seesaw2} but with the  propagators of heavy 
particles pinched and substituted by one insertion of the Weinberg operator.
This diagram is quadratically divergent
and, therefore, its contribution to
$m_{\mathrm{R}4}$
can not be reliably computed in the effective
field theory because it depends on the details of the matching with the full theory
from which the effective one originates (in fact it vanishes in dimensional regularization or 
in any other regularization scheme allowing symmetric integration), but one can use naive 
dimensional analysis to estimate contributions of order
\begin{equation}
m_{\mathrm{R}4} \sim \frac{\Lambda_W}{(4\pi)^{4}}
\sum_{ij} Y_{i4} c^*_{ij} Y_{j4} \sim
\frac{ Y_{i4} (m^*_\nu)_{ij} Y_{j4}}{(4\pi)^{4}}
\frac{\Lambda^2_W}{v^2_\phi}\, ,
\label{eq:nda-estimate}
\end{equation}
which is precisely the result obtained in the see-saw models discussed above if 
one identifies
$\Lambda_W \sim m_{\mathrm{R}k},m_\chi$. However, it is important to remark that in the
low energy effective  theory $m_{\mathrm{R}4}$ is a free parameter, and
eq.~(\ref{eq:nda-estimate}) is only a naive dimensional analysis estimate of what
one would expect in a more complete theory.

\section{Light neutrino masses induced by new generations} \label{4gto2gmasses}

After the discussion above, to describe correctly the neutrino sector
of models with four generations we need just one relatively light right-handed neutrino,
$\nu_{\mathrm{R}}$, to give Dirac mass terms to the fourth-generation neutrinos,
while Majorana masses for light neutrinos can be parametrized by the
Weinberg operator. We  will work in this minimal four-generation scenario, thus,
the relevant part of the Lagrangian for our discussion is 
\begin{equation}
\mathcal{L}_{Y} = -\bar{\ell}Y_{e} e_{\mathrm{R}} \phi -
			\bar{\ell} y \nu_{\mathrm{R}} \tilde{\phi} -
			\cfrac{1}{2} \, \overline{\nu_{\mathrm{R}}^{\mathrm{c}}} m_{\mathrm{R}}
			\nu_{\mathrm{R}} + 
			\frac{1}{2v_{\text{\ensuremath{\phi}}}^{2}} 
			(\overline{\ell}\tilde\phi)m_{\mathrm{L}}
			(\phi^{\dagger}\tilde{\ell)}+ \mathrm{H.c.}\ ,\label{eq:4geff-lagr}
\end{equation}
where $\ell$ and $e_{\mathrm{R}}$ contain the four generation components
while $\nu_{\mathrm{R}}$ is the only right-handed neutrino. Thus $Y_{e}$
is a completely general $4\times4$ complex matrix, $y$
is a $4$ component column vector, $m_{\mathrm{R}}$ is just a number and 
$m_{\mathrm{L}}$
is a general complex symmetric $4\times4$ matrix.
The Dirac limit is recovered when $m_{\mathrm{R}}=0$ and $m_{\mathrm{L}}=0$. Since
light neutrino masses are very small, we will assume $m_{\mathrm{L}}\ll v_{\phi}$
while $m_{\mathrm{R}}$, as we immediately see, cannot be very large to ensure
there are only three light active neutrinos. Moreover, as
shown in the previous section,  we do not expect it to be zero if 
$m_{\mathrm{L}}$ is not zero.

Above we have taken for $m_{\mathrm{L}}$ a general complex symmetric 
$4\times4$ matrix in spite of the fact that
to describe the light neutrino sector we just need a $3\times3$ matrix. This is 
because
in most of the neutrino mass models one also obtains contributions
to the fourth-generation Weinberg operator. For instance, we give below the values 
of
$m_{\mathrm{L}}$ one obtains for the different types of seesaw.

If the three light neutrino masses are generated by the seesaw
mechanism type I or type III, we need three of the right-handed neutrinos much
heavier than the fourth. We can always choose a basis in which the
Majorana mass matrix of right-handed neutrinos is diagonal and integrate
out the three heavy right-handed neutrinos. The result can be writen
in terms of the Weinberg operator in \eqref{eq:4geff-lagr} with
\begin{equation}
(m_{\mathrm{L}})_{\alpha\beta} = -\sum_{k=1,2,3} 
			\frac{\left(Y_{\nu}\right)_{\alpha k} \left(Y_{\nu}\right)_{\beta k}}
			{m_{\mathrm{R} k}} v_{\phi}^{2} \label{eq:mL-seesaw}\, ,
\end{equation}
where $m_{\mathrm{R} k}$ are the eigenvalues of the diagonal Majorana mass
matrix of the three heavy right-handed neutrinos, while 
$\left(Y_{\nu}\right)_{\alpha k}$
are  the Yukawa couplings 
of the three heavy right-handed neutrinos with the four lepton doublets.
Then, 
the $4\times4$ mass matrix \eqref{eq:mL-seesaw} is projective and
has at most rank 3. 

If the three light neutrino masses are generated by the VEV of a triplet
(type II see-saw), we will have
\be
(m_{\mathrm{L}})_{\alpha\beta} = 2(Y_\chi)_{\alpha\beta} v_\chi \, ,
\ee
being $(Y_\chi)_{\alpha\beta}$ the Yukawa couplings of the 4 lepton doublets
to the triplet and $v_\chi \sim \mu v_{\phi}^{2}/m_{\chi}^{2}$ its
VEV. In this case, $m_{\mathrm{L}}$ is a completely general $4\times4$ symmetric
complex matrix.

After SSB the neutrino mass matrix (in the basis 
$(\nu^\mathrm{c}_{\mathrm{L} \alpha},\nu_{\mathrm{R}})$) is 
\be
M=
\left(\begin{array}{cc}
m_{\mathrm{L}} & y v_{\phi}\\
y^{\mathrm{T}} v_{\phi} & m_{\mathrm{R}}
\end{array}\right)\, . 
\ee
To diagonalize this mass matrix we perform first a $4 \times 4$ rotation
in order to separate heavy from light degrees of freedom, so we change 
from the flavour basis ($\nu_{e},\nu_{\mu},\nu_{\tau},\nu_{E}$)
to a new basis $\nu_{\text{1}}^{\prime},\nu_{\text{2}}^{\prime},\nu_{\text{3}}^{\prime},\nu_{\text{4}}^{\prime}$
in which the first three states are light (with masses given by $m_{\mathrm{L}}$)
and only $\nu_{4}^{\prime}$ mixes with $\nu_{\mathrm{R}}$.  Then, we  have
$\nu_{\alpha}=\sum_{i}V_{\alpha i}\nu_{i}^{\prime}$ ($i=1,\text{\ensuremath{\cdots}},4$,
$\alpha=e,\mu,\tau,E$), where $V$ is a orthogonal matrix,
and we define
\begin{equation} \label{def-N}
	N_{\alpha} \equiv V_{\alpha 4} = \frac{y_\alpha}
			{\sqrt{\sum_{\beta}y_{\beta}^{2}}}\, .
\end{equation}
Now, we are free to choose $\nu_{\text{1}}^{\prime},\nu_{\text{2}}^{\prime},\nu_{\text{3}}^{\prime}$
in any combination of $\nu_{e},\nu_{\mu},\nu_{\tau},\nu_{E}$
as long as they are orthogonal to $\nu_{4}^{\prime}$, i.e., $\sum_{\alpha}
V_{\alpha k} N_{\alpha} = 0$
for $k \text{=1,2,3}$. The orthogonality of $V$ almost fixes all
its elements in terms of $N_{\alpha}$, but still leaves us some freedom
to set three of them to zero. Following \cite{Babu:1988ig,Babu:1988wk}
we choose $V_{\tau 1} = V_{E1} = V_{E2}=0$ for convenience. 
The transpose of the matrix $V$ is:
\be
\label{V}
{V^\mathrm{T}=\left(\begin{array}{cccc}
\cfrac{N_{\mu}}{\sqrt{N_{e}^{2}+N_{\mu}^{2}}} & \cfrac{-N_{e}}{\sqrt{N_{e}^{2}+N_{\mu}^{2}}} & 0 & 0\\
\cfrac{N_{e}N_{\tau}}{\sqrt{(N_{e}^{2}+N_{\mu}^{2})(1-N_{E}^{2})}} & 
\cfrac{N_{\mu}N_{\tau}}{\sqrt{(N_{e}^{2}+N_{\mu}^{2})(1-N_{E}^{2})}} & 
\cfrac{-N_{e}^{2}-N_{\mu}^{2}}{\sqrt{(N_{e}^{2}+N_{\mu}^{2})(1-N_{E}^{2})}} & 0\\
\cfrac{N_{e}N_{E}}{\sqrt{(1-N_{E}^{2})}} & \cfrac{N_{\mu}N_{E}}{\sqrt{(1-N_{E}^{2})}} & 
\cfrac{N_{\tau}N_{E}}{\sqrt{(1-N_{E}^{2})}} & - {\sqrt{(1-N_{E}^{2})}}\\
N_{e} & N_{\mu} & N_{\tau} & N_{E}\, .
\end{array}\right)}
\ee
After this rotation the neutrino mass matrix is
\be
\label{Mtilde}
\tilde{M}=
\left(\begin{array}{cc}
\tilde{m}_{\mathrm{L}} & \begin{array}{cc}
\omega_{1} & 0\\
\omega_{2} & 0\\
\omega_{3} & 0
\end{array}\\
\begin{array}{ccc}
\omega_{1} & \omega_{2} & \omega_{3}\\
0 & 0 & 0
\end{array} & \begin{array}{cc}
\omega_{4} & m_{\mathrm{D}}\\
m_{\mathrm{D}} & m_{\mathrm{R}}
\end{array}
\end{array}\right)\, ,
\ee
where $(\tilde{m}_{\mathrm{L}})_{k k^\prime}=(V m_{\mathrm{L}} 
V^{\mathrm{T}})_{k k^\prime}$ is a $3\times3$
matrix with $k,k^\prime=1,2,3$, $\text{\ensuremath{\omega}}_{k}=
(V m_{\mathrm{L}} V^{\mathrm{T}})_{4k}$, $\omega_{4}=
(Vm_{\mathrm{L}}V^{\mathrm{T}})_{44}$
and $m_{\mathrm{D}}=v_\phi \sqrt{\sum_{\alpha}y_{\alpha}^{2}}$. 
Since $\tilde{m}_{\mathrm{L}},\omega_k,\omega_4 \ll m_{\mathrm{R}},
m_{\mathrm{D}}$,
the matrix  $\tilde{M}$ can be block-diagonalized using the see-saw formula. 
Then, the mass matrix
of the light neutrinos (at tree level) will be 
\begin{equation}
m_{\nu}^{(0)}=\tilde{m}_{\mathrm{L}} - 
		\frac{m_{\mathrm{R}}}{m_{\mathrm{R}} \omega_{4} - 
		m_{\mathrm{D}}^{2}} \vec{\omega} \cdot \vec{\omega}^{\mathrm{T}}
		\label{eq:mlight0}\, ,
\end{equation}
while the heavy sector will be obtained after diagonalizing the $2\times2$
matrix 
\be
\label{MH}
M_H = 
\left(\begin{array}{cc}
\omega_{4} & m_{\mathrm{D}}\\
m_{\mathrm{D}} & m_{\mathrm{R}}
\end{array}\right)\ . 
\ee
Neglecting $\omega_4$, this diagonalization leads to two Majorana neutrinos
\bea
\nu_{4}&=&i\cos\theta(-\nu_{4}^{\prime}+\nu_{4}^{\prime \mathrm{c}})
		+i \sin\theta(\nu_{\mathrm{R}}-\nu_{\mathrm{R}}^{\mathrm{c}})
\label{eq:mixingsfourth1}\\
\nu_{\bar{4}}&=&-\sin\theta(\nu_{4}^{\prime}+\nu_{4}^{\prime \mathrm{c}})
		+\cos\theta(\nu_{\mathrm{R}}+\nu_{\mathrm{R}}^{\mathrm{c}})
\label{eq:mixingsfourth2}
\eea
with masses
\begin{equation}
m_{4,\bar{4}}=\cfrac{1}{2}\left(\sqrt{m_{\mathrm{R}}{}^{2}
		+ 4m_{\mathrm{D}}^{2}}\mp m_{\mathrm{R}}\right)  \ ,  
		\label{eq:massesfourth}
\end{equation}
and mixing angle $\tan^{2}\theta=m_{4}/m_{\bar{4}}$.
The imaginary unit factor $i$ and the relative signs in $\nu_{4}$ are necessary
to keep the mass terms positive and preserve the canonical Majorana
condition $\nu_{4}=\nu_{4}^{\mathrm{c}}$. If $m_{\mathrm{R}}\ll 
m_{\mathrm{D}}$, we have $m_{4}\approx m_{\bar{4}}$,
$\tan\theta\approx1$, and we say we are in the pseudo-Dirac limit
 while when $m_{\mathrm{R}}\gg m_{\mathrm{D}}$, $m_{4}\approx 
 m_{\mathrm{D}}^{2}/m_{\mathrm{R}}$ and
$m_{\bar{4}}\approx m_{\mathrm{R}}$, $\tan\theta\approx 
m_{\mathrm{D}}/m_{\mathrm{R}}$ and
we say we are in the see-saw limit.

Eq.~\eqref{eq:mlight0} can be used as long as $m_{\mathrm{R}}\omega_{4}
-m_{\mathrm{D}}^{2}$
is different from zero. However, we expect $m_{\mathrm{R}}$ to be below few
TeV and $\omega_{4}$ below $1$~eV. Therefore $m_{\mathrm{R}}\omega_{4}\ll 
m_{\mathrm{D}}^{2}$
unless $m_{\mathrm{D}}$ is very small but, in that case, the fourth-generation
neutrinos will be too light. Thus, the correction to the $3\times3$
neutrino mass matrix is projective (only one eigenvalue different
from zero) and it is naturally order $m_{\mathrm{L}}^{2}$ and, 
therefore, negligible.

Summarizing, there are two heavy neutrinos $4$ and $\bar{4}$ (with a small
pollution from $m_{\mathrm{L}}$ which can be neglected) and a 
tree-level mass matrix for the light neutrinos 
$m^{(0)}_\nu \simeq \tilde{m}_{\mathrm{L}}$. Therefore,
 neglecting the  small $\omega_i$'s in eq.~\eqref{Mtilde}, the 5$\times$5 
 unitary matrix which relates the flavour 
 with the mass eigenstate basis can be written as $U = U_H \cdot U_L$, being 
 $U_H$ the rotation in the heavy sector which diagonalizes 
 the mass matrix $M_H$ in eq.~\eqref{MH} and $U_L$
given by 
\be
U_L=
\left(\begin{array}{cc}
V & \begin{array}{c}
0\\
 0\\
 0\\
 0
\end{array}\\
\begin{array}{cccc}
0 & 0 & 0 &0
\end{array} 
& \begin{array}{c}
1
\end{array}
\end{array}\right)  
\left(\begin{array}{cc}
W & \begin{array}{cc}
0 & 0\\
0 & 0\\
0 & 0
\end{array}\\
\begin{array}{ccc}
0 & 0 & 0 \\
0 & 0 & 0
\end{array} & \begin{array}{cc}
1 & 0\\
0 & 1
\end{array}
\end{array}\right)\, ,
\ee
where $V$ rotates from the $\nu_i'$ basis to the flavour basis (see eq.~\eqref{V})
and $W$ is the matrix which diagonalizes  $\tilde{m}_{\mathrm{L}}$. 
Within this approximation,  the mixing among the light and the 
heavy sector, which we wish to constrain, depends on 
$(U_L)_{\alpha 4} = V_{\alpha 4} = N_\alpha$.  

Having fixed the tree-level neutrino mass spectrum and given the huge
hierarchies present we should consider the stability of the results
against radiative corrections. 
One can  check that there are no rank-changing one-loop corrections
to the neutrino mass matrices. This result can be easily  understood in the $\nu'_i$ basis 
that we defined before, since the light neutrinos 
$(\nu_{\text{1}}^{\prime},\nu_{\text{2}}^{\prime},\nu_{\text{3}}^{\prime})$
are decoupled from the heavy sector, $\nu_{4}, \nu_{\bar{4}}$,
so there are not one-loop diagrams involving the fourth-generation neutrinos  
with light ones as external legs.

\FIGURE{
\begin{centering}
\includegraphics[width=0.6\textwidth]{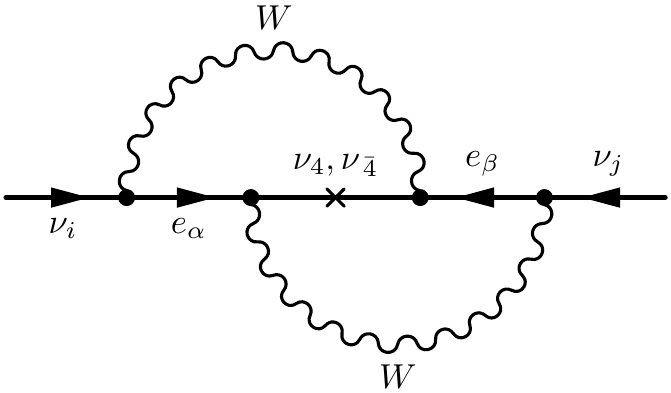}
\end{centering}
\caption{Two-loop diagram generating light neutrino masses in the presence 
of a Majorana fourth generation. \label{fig:light_2loops}}
}

However it has been shown \cite{Petcov:1984nz,Babu:1988ig,Babu:1988wk,Aparici:2011nu} 
that two-loop corrections induced by the
fourth-generation fermions can generate neutrino masses for the light
neutrinos even if they were not present at tree level, see figure \ref{fig:light_2loops}. 
In the $\nu'_i$ basis  the result
reads (see \cite{Aparici:2011nu} for details) 
\be \label{2-loop-masses}
(m_{\nu})_{ij}^{(2)}=-\frac{g^{4}}{m_{W}^{4}} m_{\mathrm{R}}
		m_{\mathrm{D}}^{2}\sum_{\text{\ensuremath{\alpha}}}
V_{\alpha i}V_{\alpha4}m_{\alpha}^{2}\sum_{\beta}V_{\beta j}V_{\beta4}m_{\beta}^{2}I_{\alpha\beta}\, ,
\ee
where the sums run over the charged leptons $\alpha,\beta=e,\mu,\tau,E$
while $i,j=1,2,3$, and $I_{\alpha\beta}$ is a loop integral which
was discussed in \cite{Aparici:2011nu}. When 
$m_{\mathrm{R}}=0,$ $(m_{\nu})_{ij}^{(2)}=0$,
as it should, because in that case lepton number is conserved. Also
when $m_{\mathrm{D}}=0$ we obtain $(m_{\nu})_{ij}^{(2)}=0$, since then the
right-handed neutrino decouples completely and lepton number is again
conserved.

To see more clearly the structure of this mass matrix we can approximate 
 $m_{e}=m_{\mu}=m_{\tau}=0$; then, since
we have chosen $V_{\tau1}=V_{E1}=V_{E2}=0$, the only non-vanishing
element in $(m_{\nu})_{ij}^{(2)}$ is $(m_{\nu})_{33}^{(2)}$ and
it is proportional to $V_{E3}^{2}N_{E}^{2}m_{E}^{4}I_{EE}$. 
Therefore,
the largest  contribution  to $(m_\nu)^{(2)}$ is given by: 
\bea
(m_\nu)_{33}^{(2)} &=& - \frac{g^4}{m_W^4} N_E^2 (N_e^2 +N_\mu^2 + N_\tau^2)
m_{\mathrm{R}} m_{\mathrm{D}}^2 m_E^4 I_{EE}
\nonumber \\
& \approx &  \frac{g^4}{2 (4\pi)^4}(N_e^2 +N_\mu^2 + N_\tau^2)
m_{\mathrm{R}} \, \frac{m_{\mathrm{D}}^2 m_E^2} {m_W^4}  
\ln \frac{m_E}{m_{\bar{4}}} \ ,
\label{eq:m33}
\eea
where in the last line  we have used the approximated expression of the loop integral 
$I_{EE}$ in the case $m_E \gg m_{4,\bar{4}} \gg m_W$ for definiteness, 
but other mass relations lead to analogous conclusions. 
Keeping all the charged lepton masses 
one can easily show that the eigenvalues of the light
neutrino mass matrix are proportional to $m_{\mu}^{4},\, m_{\tau}^{4},\, m_{E}^{4}$
which gives a huge hierarchy between neutrino masses. Therefore, as
discussed in \cite{Aparici:2011nu,Aparici:2011eb}, these radiative corrections cannot explain
by themselves the observed spectrum of masses and mixings, although they 
lead to a strong constraint for this kind of SM extensions which we will 
analyze in the next section.

\section{Phenomenological constraints} \label{pheno}

\subsection{Direct searches}

\FIGURE{
	\centering
	\includegraphics[width=0.8\textwidth]{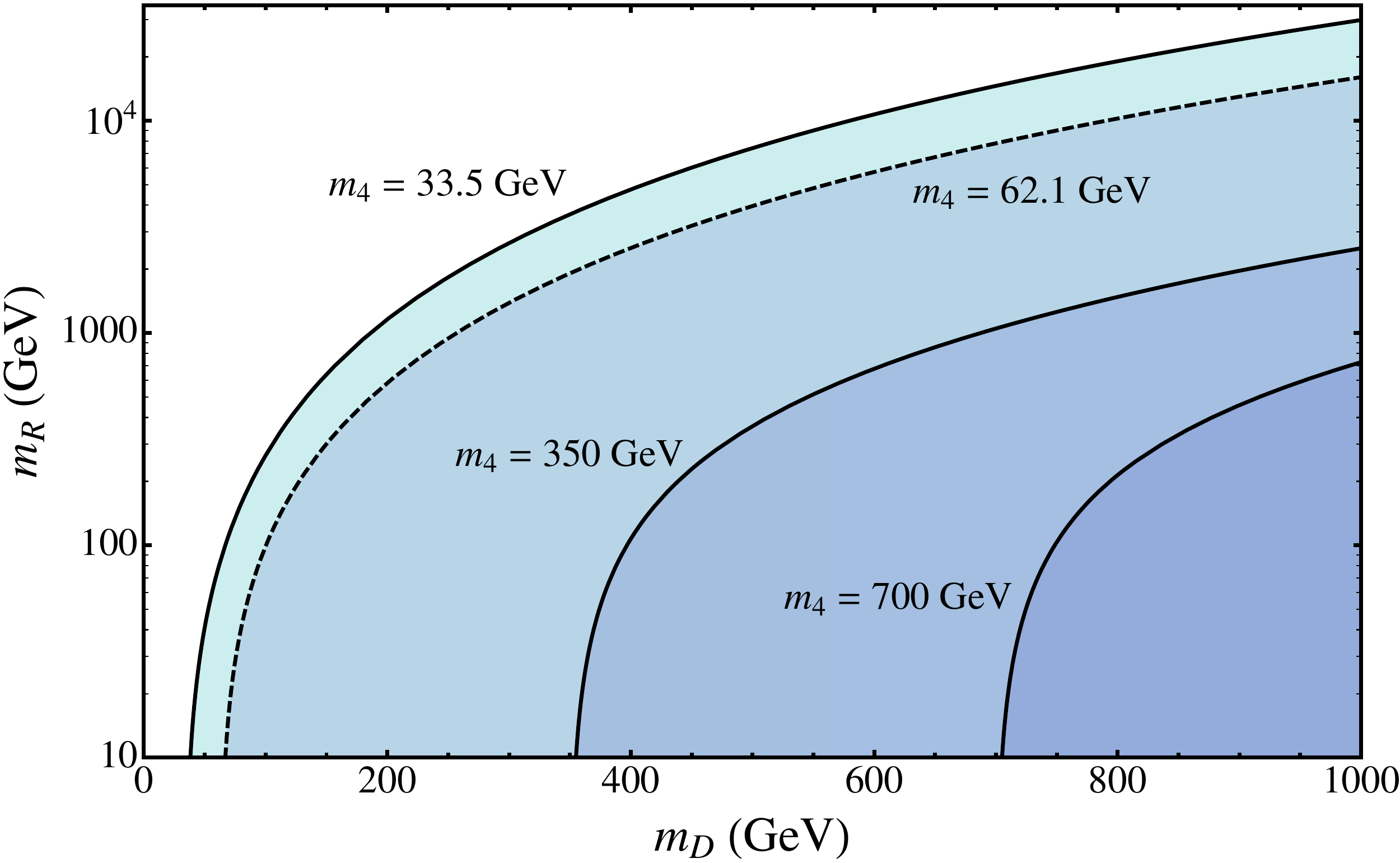}
	\caption{The shaded region shows the allowed values for the Majorana 
			and Dirac masses of	the heavy neutrinos given the LEP bound 
			$m_N > 33.5 \; \mathrm{GeV}$ on stable neutrinos with
			both Dirac and Majorana masses. We also display a dashed
			line in the 62.1 GeV limit for unstable neutrinos.
			Two more lines are drawn for completeness, giving an idea 
			of the combination of parameters that produces two possible,
			allowed masses for the lightest heavy neutrino.
			} \label{grafmN}
}

Let us now discuss the constraints that several phenomenological tests
impose on the parameters of this minimal four-generation (4G) model. Direct searches
for the new heavy leptons can be used to set limits on the Yukawa 
couplings and the Majorana mass of the $\nu_\mathrm{R}$. In the case 
of the heavy charged lepton, searches at LEP \cite{Nakamura:2010zzi} yield 
$m_E > 100.8 \; 
\mathrm{GeV}$ (assuming it decays rapidly to $\nu W$; a slightly poorer
bound is obtained if the lepton is long-lived and can be tracked inside
the detectors), which can be immediately translated into a bound on the 
corresponding Yukawa. For the heavy neutrinos, we can have different
bounds depending on their stability and the Dirac or Majorana
character of their masses \cite{Nakamura:2010zzi}: stable neutrinos, 
for example (understood here
as `stable enough to get out of the detectors after production'), 
are only constrained
by the requirement that they don't show up in the invisible decays
of the $Z$ boson. Unstable (visible) neutrinos get tighter bounds due
to the non-observation of their decay products. As we
are not making any \emph{a priori} assumption about the neutrino mass
structure, we will select here the most conservative from this set of bounds;
that corresponds to a stable neutrino with both Dirac and Majorana
mass terms, for which we
demand $m_N > 33.5 \; \mathrm{GeV}$ \cite{Carpenter:2010sm}. 
The weakest
bound for an unstable neutrino, which applies if it has again both
Dirac and Majorana mass terms, will also be of use; we need in that case
$m_N > 62.1 \; \mathrm{GeV}$ \cite{Carpenter:2010dt}. 
As these bounds apply to 
the physical masses of the neutrinos, which as seen in 
eqs.~(\ref{eq:mixingsfourth1}--\ref{eq:massesfourth}) are nonlinear combinations of the
Dirac and Majorana components, we display in figure \ref{grafmN}
the translation of the 33.5 and 62.1 GeV bounds into the
$m_\mathrm{D} - m_\mathrm{R}$ plane, together with several other lines
to give an idea of the relations between physical masses and Lagrangian
parameters.

As explained in section \ref{4gto2gmasses}, the neutrino Yukawas $y_\alpha$
encode the mixings between the flavour-eigenstate neutrinos $\nu_\alpha$
and the mass eigenstates $\nu_{4, \bar 4}$. Thus, we can use 
mixing-mediated LFV processes to constrain the values of
the light neutrino Yukawas $y_e$, $y_\mu$, $y_\tau$. It is important
to note, however,
that the situation is not the same for `stable' and unstable neutrinos; 
so-called stable neutral leptons are constrained to decay outside the 
detectors, which implies that the mean free path must go beyond 
$\mathcal{O} (\mathrm{m})$. The lightest of our heavy neutrinos can
only decay through mixing (the main channel being $\nu_4 \rightarrow 
\ell_\alpha W$, $\alpha=e,\mu,\tau$, with a possibly virtual $W$ depending on the mass
of the $\nu_4$), so this statement is actually a constraint
on the Yukawas, implying $y_\alpha \sim N_\alpha \lesssim 10^{-6}$. This
constraint is much stronger than any other phenomenolgical bound, and
so it ends the discussion for stable neutrinos, which must
have very small mixings that won't be observable in low-energy
experiments in the near future (see below). For the rest of this section 
we will consider the case of unstable neutrinos,
which present a richer variety of constraints.

\subsection{Lepton flavour violation}

Let us now discuss the bounds on violation of lepton family number 
that can shed light on the relevant mixings of our model; the most 
stringent limits are derived from the non-observation of radiative decays
of the form $\ell_\alpha \rightarrow \ell_\beta \gamma$~\footnote{Bounds obtained from
present data on $\mu$--$e$ conversion in nuclei~\cite{Nakamura:2010zzi} are of the same order.
However, there are plans to improve the sensitivity in $\mu$--$e$
conversion in $4$ and even $6$ orders of magnitude \cite{Hungerford:2009zz},
therefore we expect from this process much stronger bounds in the future.}. In our model,
the ratios for such processes are given by
\begin{equation} \label{radiative-rates}
	B (\ell_\alpha \rightarrow \ell_\beta \gamma) \equiv 
		\frac{\Gamma (\ell_\alpha \rightarrow \ell_\beta \gamma)}
			{\Gamma (\ell_\alpha \rightarrow \ell_\beta \nu \bar \nu)}
		= \frac{3 \alpha}{2 \pi} \, \left| \sum_{a=4,\bar{4}} U_{\beta a} 
			U^\ast_{\alpha a} \, H \left( \nicefrac{m_a^2}{m_W^2}
			\right) \right|^2\, ,
\end{equation}
where $H(x)$ is a loop function that can be found in \cite{Buras:2010cp},
and the sum proceeds over all the heavy neutrinos\footnote{Note this 
expression contains the contributions from the light neutrinos; by using
unitarity of the mixing matrix, they are included in the definition of $H(x)$.} 
(one in the Dirac case,
two if they are Majorana). The weakest bounds are obtained if only one
neutrino with light mass runs inside the loop; this corresponds either to
the Dirac limit with a low mass or to a hard see-saw limit, with 
the heavy neutrino almost decoupled due to its small mixing. We will assume
this scenario in our calculations in order to produce conservative bounds.
Table \ref{table-radiative} summarises the experimental limits and the
constraints that can be extracted from these processes.

\TABLE{ 
	\centering
	\begin{tabular}{| c | c |}
		\hline
		Experimental bounds at 90\% C.L. \cite{Nakamura:2010zzi,Adam:2011ch}
			&	Constraints on the mixings
		\\ \hline
		$B (\mu \rightarrow e \gamma) < 2.4 \times 10^{-12}$ &
			$N_e N_\mu < 2.85 \times 10^{-4}$
		\\ \hline
		$B (\tau \rightarrow e \gamma) < 1.85 \times 10^{-7}$ &
			$N_e N_\tau < 0.079$
		\\ \hline
		$B(\tau \rightarrow \mu \gamma) < 2.5 \times 10^{-7}$ &
			$N_\mu N_\tau < 0.093$
		\\ \hline
	\end{tabular}
	\caption{Summary of the constraints derived from low-energy
		radiative decays.}
		\label{table-radiative}
}

\subsection{Universality tests}

A second class of constraints upon family mixing arises from the tests
of universality in weak interactions. For our purposes, these 
are either direct comparison
of decay rates of one particle into two different weak-mediated channels, or 
comparison of the decay rates of two different particles into the same
channel~\footnote{Data from neutrino oscillations
can also be used to constrain the elements of the leptonic mixing matrix \cite{Antusch:2006vwa}, however, 
they lead to weaker bounds than the ones obtained here.}. If the weak couplings are to be the same for all families
these rates should differ only in known kinematic factors or calculable 
higher-order corrections. The relevant ratios are:
\begin{align*}
	R_{\pi \rightarrow e / \pi \rightarrow \mu} &\equiv 
		\frac{\Gamma (\pi \rightarrow e \nu) }
			{\Gamma (\pi \rightarrow \mu \nu) } 
	\, , &  R_{\tau \rightarrow \mu / \tau \rightarrow e}
		&\equiv \frac{\Gamma (\tau \rightarrow \mu \nu \bar \nu)}
			{\Gamma (\tau \rightarrow e \nu \bar \nu) }
	\, , \\
	R_{\tau \rightarrow e / \mu \rightarrow e} &\equiv 
		\frac{\Gamma(\tau \rightarrow e \nu \bar \nu)}
			{\Gamma (\mu \rightarrow e \nu \bar \nu)}
	\, ,  &  R_{\tau \rightarrow \mu / \mu \rightarrow e}
		&\equiv \frac{\Gamma (\tau \rightarrow \mu \nu \bar \nu)}
			{\Gamma (\mu \rightarrow e \nu \bar \nu)} \, , 
\end{align*}
and their theoretical values in a 3G SM can be consulted, for example, in
\cite{Pich:1997hj}. Comparison of the experimental values and the 3G
predictions yields values very close to 1, as can be seen in table 
\ref{table-universality}; in our 4G model, family mixing induces deviations
from this behaviour that must be kept under control. Essentially, these
deviations result from the fact that the flavour-eigenstate neutrinos
$\nu_e, \nu_\mu, \nu_\tau$ have a small component of the heavy neutrinos
$\nu_4, \nu_{\bar 4}$, which cannot be produced in the decays of pions,
taus or muons; the corresponding mixings $N_e, N_\mu, N_\tau$ are then
forced to be small. In table \ref{table-universality} we also show
the constraints that this processes impose on the mixing
parameters.
\TABLE{ 
	\centering
	\begin{tabular}{| c | c |}
		\hline
		Experimental bounds at 90\% C.L. \cite{Nakamura:2010zzi}
			&	Constraints on the mixings
		\\ \hline
		$\cfrac{R_{\pi \rightarrow e / \pi \rightarrow \mu}}
			{R_{\pi \rightarrow e / \pi \rightarrow \mu }^\mathrm{SM}}
			= 0.996 \pm 0.005$  &
		$N_e^2 - N_\mu^2 = 0.004 \pm 0.005$
		\\ \hline
		$\cfrac{R_{\tau \rightarrow e / \tau \rightarrow \mu}}
			{R_{\tau \rightarrow e / \tau \rightarrow \mu }^\mathrm{SM}}
			= 1.000 \pm 0.007$  &
		$N_e^2 - N_\mu^2 = 0.000 \pm 0.007$
		\\ \hline
		$\cfrac{R_{\tau \rightarrow e / \mu \rightarrow e}}
			{R_{\tau \rightarrow e / \mu \rightarrow e }^\mathrm{SM}}
			= 1.003 \pm 0.007$  &
		$N_\mu^2 - N_\tau^2 = 0.003 \pm 0.007$
		\\ \hline
		$\cfrac{R_{\tau \rightarrow \mu / \mu \rightarrow e}}
			{R_{\tau \rightarrow \mu / \mu \rightarrow e }^\mathrm{SM}}
			= 1.001 \pm 0.007$  &
		$N_e^2 - N_\tau^2 = 0.001 \pm 0.007$
		\\ \hline
	\end{tabular}
	\caption{Summary of the constraints derived from universality tests
		in weak decays. The ratios marked as ``SM'' represent the 
		theoretical predictions of a 3G Standard Model.}
		\label{table-universality}
}

\FIGURE{
	\centering
	\includegraphics[width=\textwidth]{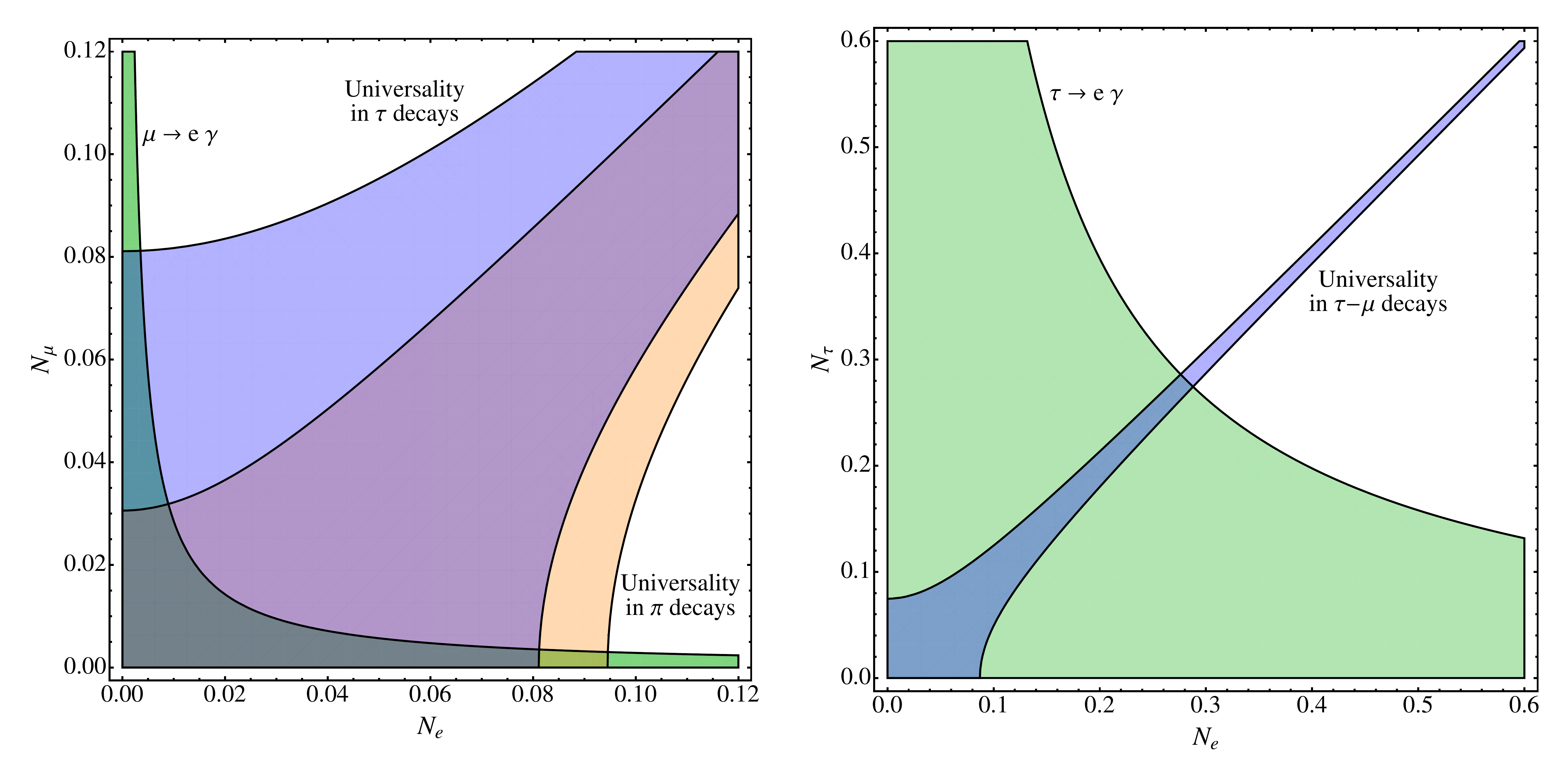}
	\caption{These two graphs present the allowed regions for the mixing
		parameters in our model at 90\% confidence level, 
		according to several LFV tests. The left
		plot displays the constraints in the $N_e - N_\mu$ plane, which
		are much more stringent and suffice to bound both $N_e$ and $N_\mu$.
		The right plot displays the $N_e - N_\tau$ plane; the $N_\mu - 
		N_\tau$ plane offers slighly poorer constraints and is not
		displayed.}
	\label{grafLFV}
}

In figure \ref{grafLFV} we collect all the relevant LFV constraints from
tables \ref{table-radiative} and \ref{table-universality}. As can be 
read from the graphs, the final bounds we can set on the mixings of the
light families are
\begin{align}
	N_e    &< 0.08 \nonumber \\
	N_\mu  &< 0.03 \label{LFV-bounds} \\
	N_\tau &< 0.3  \nonumber
\end{align}

\FIGURE{
	\centering
	\includegraphics[width=0.9\textwidth]{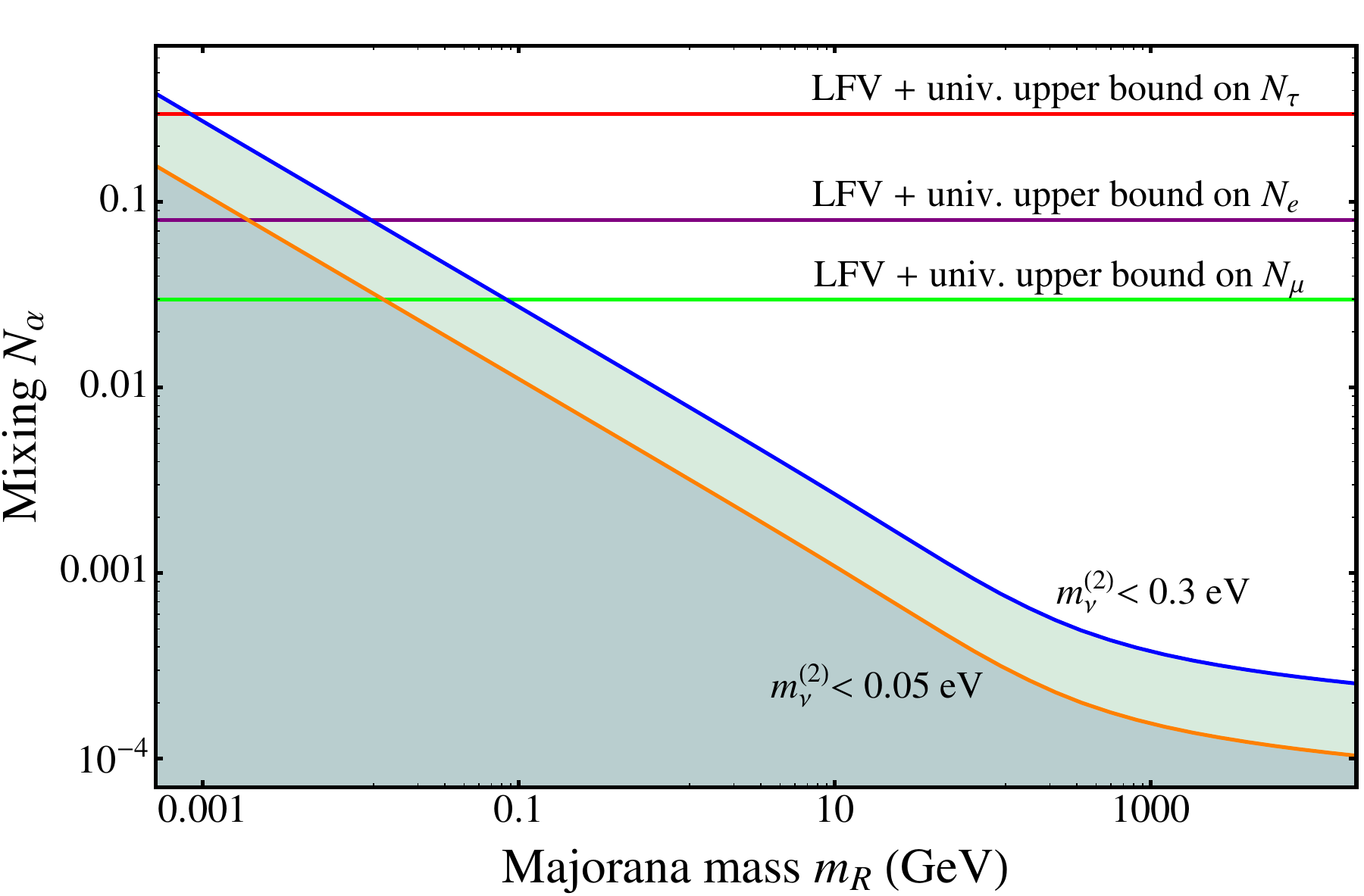}
	\caption{Summary of the constraints on the mixings of the model,
		as defined in equation \eqref{def-N}. The three horizontal lines
		present the upper bounds in equation \eqref{LFV-bounds},
		derived from universality tests and limits on LFV processes.
		The two shaded areas display the allowed region derived from
		the fact that the two-loop diagram does not disturb the
		correct structure for the light neutrino masses, assumed to
		arise from any other mechanism. This last bound applies to
		any of the mixings.
	} \label{fig:cosmo}
}

\subsection{Light neutrino masses}

Finally, there is a further constraint that can be set upon the mixings
of the model: as explained in section \ref{4gto2gmasses},
the two-loop mechanism which gives small Majorana masses for the light 
neutrinos cannot explain by itself the observed pattern of masses in this 
simple model; it, nevertheless, still has the potential to generate 
\emph{too large}
masses, which would exclude the model. Of course, one could always invoke 
cancellations between these two-loop masses and other contributions
(for example, the Weinberg operator), but we think that this wouldn't be 
a natural situation and choose not to consider it. If we bar such 
cancellations we need to impose that the two-loop masses don't go
above some limit, and thus a bound can be set 
upon the parameters that participate in the two-loop mechanism, essentially
the mixings and the Majorana mass, as seen in equation~\eqref{2-loop-masses}.
Figure~\ref{fig:cosmo} shows the allowed regions for this constraint;
the curves are constructed using the lowest possible values of the 
fourth-generation Dirac  masses, in order to provide conservative limits
(this implies using a different value of $m_{\mathrm{D}}$ for each
$m_{\mathrm{R}}$, as
we must also impose that $m_4$ is above 62.1 GeV). 
We show two possible limits: $m_\nu^{(2)} < 0.05 \; \mathrm{eV}$
ensures that the largest two-loop mass is below the atmospheric mass scale;
this, of course, does not guarantee that it 
doesn't distort the neutrino spectrum, which may contain smaller
masses, so this 
bound can be contemplated as rather conservative (particular models may
need to impose a stronger bound to be phenomenologically viable).
An even more conservative constraint
is obtained if we impose $m_\nu^{(2)} < 0.3 \; \mathrm{eV}$, meaning
that the largest of the two-loop masses is not above the bound imposed
by cosmology to each mass of the degenerate spectrum, 
$\sum_k m_k \lesssim 1$ eV \cite{Hannestad:2010yi,GonzalezGarcia:2010un}.
Two-loop masses
as large as 0.3 eV will in most cases spoil the structure of neutrino masses,
but there may be pathological cases in which such situation is allowed
(for example, if the Weinberg operator generates a massless neutrino and 
two massive ones near the 0.3 eV limit; then the two-loop diagram might provide
the third mass to fit the mass splittings). Even with these conservative
assumptions, the two-loop bound proves to be much stronger than those 
derived from universality and LFV for most of the parameter space. It is,
therefore, a limit to be kept in mind when considering 4G models with
Majorana neutrinos.

\subsection{Neutrinoless double beta decay ($0\nu 2\beta$)}

In our framework, the contributions to the amplitude of 
neutrinoless double beta decay
($0\nu 2\beta$) can be written as:
\be
A = A_L +  A_{\rm{md}} + A_4   \ , 
\ee
where $A_L$ stands for the light neutrino contribution 
(i.e., neutrino masses $m_k \ll p_{\rm{eff}}  \sim 100\,$ MeV), given by 
\be
\label{AL}
A_L \propto \sum_k^{\rm {light}} m_k U_{e k}^2 M^{0\nu 2\beta}(m_k)  
\simeq  m_{ee} M^{0\nu 2\beta}(0) \ ,
\ee
with $M^{0\nu 2\beta}(0)\propto 1/p_{\rm{eff}} ^2$ the nuclear matrix element.
The cosmology upper bound on the sum of neutrino masses, 
$\sum_k m_k \lesssim  1$ eV \cite{Hannestad:2010yi,GonzalezGarcia:2010un},
combined with neutrino oscillation data, leads to an upper limit on
each neutrino mass  $m_k \lesssim 0.3$ eV and on 
the element of the neutrino mass matrix relevant to $0\nu 2\beta$ decay,
$m_{ee} \lesssim 0.3$ eV.
 
 $A_{\rm{md}}$ represents the 
additional, model dependent contribution due to the unknown physics which 
generates the three light neutrino masses parametrized by the Weinberg 
operator. We assume that this last term is negligible compared to 
$A_L$, as it is the case 
if the underlying mechanism for neutrino masses is any of the standard 
three see-saw types \cite{Blennow:2010th}.

We focus then on the contribution from the fourth-generation neutrinos 
($\nu_4,\nu_{\bar{4}}$),  
given by
\bea 
A_4 &\propto& N_{e}^2 \left(
m_4 \cos^2 \theta M^{0\nu 2\beta}(m_4)  
 -  m_{\bar{4}} \sin^2 \theta  M^{0\nu 2\beta}(m_{\bar{4}})   \right)  
 \nonumber \\
\label{A4}
& \propto& N_{e}^2
 \left( \frac{\cos^2 \theta}{m_4} - \frac{\sin^2 \theta}{m_{\bar{4}}} \right)  
= N_{e}^2  \frac{m_{\bar{4}}^2 - m_4^2}{m_4 m_{\bar{4}} (m_4 + m_{\bar{4}})}
= N_{e}^2  \frac{m_{\mathrm{R}}}{m_{\mathrm{D}}^2}\, , 
\eea
where we have used that 
$M^{0\nu 2\beta}(m_a) \propto  1/m_a^2$ for $a=4,\bar{4}$,
$\tan^2\theta = m_4/m_{\bar{4}}$,  
$m_4 m_{\bar{4}} = m_{\mathrm{D}}^2$ and 
$m_{\bar{4}} - m_4 = m_{\mathrm{R}}$. 
From eqs.~(\ref{AL}) and  (\ref{A4}) we see that 
 the fourth-generation neutrino contribution to the $0\nu 2\beta$ amplitude 
 can be dominant provided $N_e^2 m_{\mathrm{R}}/m_{\mathrm{D}}^2 
 >   m_{ee} / (100 \, {\rm MeV})^{2}$.
Notice, in fact, that the value of
$m_{ee}$ could be zero if normal hierarchy is realised and the neutrino
phases have the appropriate values; in this extreme case the only contribution
would be that of the fourth generation, which would dominate $0 \nu 2 \beta$.

Now we can exploit the dependence on $N_e^2 m_{\mathrm{R}}$  of both 
$A_4$ and $(m_\nu)_{33}^{(2)}$ in eq.~\eqref{eq:m33}
to constrain the fourth-generation neutrino contribution to the 
$0\nu 2\beta$ decay amplitude
\footnote{Note that  
$(m_\nu)_{33}^{(2)}$ receives contributions from $N_e$, $N_\mu$ and $N_\tau$,
while $0 \nu 2 \beta$ only involves $N_e$.}
, namely 
\be \label{A4-bound}
A_4 \leq  \left( \frac {4\pi m_W}{g m_{\mathrm{D}}} \right)^4 
\frac{2 (m_\nu)_{33}^{(2)}}{m_E^2   \ln \frac{m_E}{m_{\bar{4}}}} 
\lesssim 190 (m_\nu)_{33}^{(2)} \left( \frac{50 \,  {\rm GeV}}
{m_{\mathrm{D}}} \right)^4 \, {\rm GeV}^{-1} \, , 
\ee
where we have taken into account the LEP limit, $m_E \gtrsim 100$ GeV
and set $\ln(m_E/m_{\bar{4}}) \simeq1$. 
From this equation, it is clear that the largest 
fourth-generation contributions to the amplitude $A_4$ correspond to 
a small Dirac neutrino mass, $m_{\mathrm{D}}$. 
Imposing that the two-loop mass matrix element 
$(m_\nu)_{33}^{(2)}$ is below the cosmology upper bound, 0.3 eV,
we obtain $A_4 < 6 \times 10^{-8} (50 \, {\rm GeV}/m_{\mathrm{D}})^4 
\, {\rm GeV}^{-1}$, 
while if we require that the two-loop contribution is at most 
the atmospheric mass scale, 0.05 eV, we find  
$A_4 <  10^{-8} (50 \, {\rm GeV}/m_{\mathrm{D}})^4 \, {\rm GeV}^{-1}$. 
On the other hand, the non-observation of $0 \nu 2 \beta$
implies that $A_4 <  10^{-8}  \, {\rm GeV}^{-1}$ \cite{Pas:2000vn},
while future sensitivity is expected to improve this limit one order of
magnitude. Bringing these two results together, we see that,
once the constraint from light neutrino masses 
is taken into account, the contribution of the fourth-generation neutrinos to 
the $0 \nu 2 \beta$ decay amplitude can reach observable values 
only in the small region of parameter space 
$m_{\mathrm{D}} \lesssim 100 \; \mathrm{GeV}$ 
(see figure \ref{grafmN}), even though it is the dominant one for a larger  set
of allowed masses and mixings.

\subsection{Four generations and the Higgs boson}

It is well known that due to the presence of a new generation  there
is an enhancement of the Higgs-gluon-gluon vertex, which arises from
a triangle diagram with all quarks running in the loop. This vertex
is enhanced approximately by a factor 3 in the presence of a heavy
fourth generation, therefore the Higgs production cross section through
gluon fusion at the LHC is enhanced by a factor of 9. 
However, Higgs decay
channels are also strongly modified, in particular the Higgs to
gluon decays are equally enhanced, while the $\gamma\gamma$ channel is reduced
because of a cancellation between the quark and $W$ contributions.
Moreover some of these channels, $\gamma\gamma$ for instance, suffer from important electroweak radiative corrections \cite{Denner:2011vt}. 

With the first LHC data, ATLAS and CMS ruled out at $95\%$
C.L the range $120-600$ GeV for a SM4 Higgs boson, assuming very
large masses for the fourth-generation particles \cite{ATLAS-CMS-CONF-2011-157}.
However, different authors have noticed that if fourth-generation
neutrinos are light enough ($m_{W}/2\lesssim m_{\nu_{4}}\lesssim m_{W}$),
the decay mode of the Higgs into fourth-generation neutrinos can be
dominant for $m_H \lesssim 2 m_W$ \cite{Belotsky:2002ym,Keung:2011zc,Carpenter:2011cm,Cetin:2011fp}.
Moreover, if the lightest fourth-generation neutrino is long-lived
this decay channel is invisible and 
the excluded range for the SM4 Higgs boson is reduced
to $160-500$ GeV~\cite{Carpenter:2011cm,Cetin:2011fp}.
In general, if the  fourth-generation neutrinos have both Dirac 
($m_{\mathrm{D}}$) and Majorana ($m_{\mathrm{R}}$)
masses, the Higgs can decay to more channels: $\nu_{4}\nu_{4}$,
$\nu_{\bar{4}}\nu_{\bar{4}}$ and $\nu_{4}\nu_{\bar{4}}$.

Recently, ATLAS and CMS have analysed new data, including more 
Higgs decay channels \cite{ATLAS:2012ae,Chatrchyan:2012tx}, and they
have a preliminary low-mass ($\sim125$ GeV) hint of the Higgs boson
in several channels\footnote{Also Fermilab CDF and D0 have presented some preliminary results pointing to some excess around this mass which can 
be assigned mainly to $H\rightarrow b\bar{b}$ decays in $HW$ and $HZ$ associated production~\cite{TEVNPH:2012aa}.}. 
In particular there is an excess in the $\gamma\gamma$
channel with respect to the SM3 prediction.  
For such a light Higgs,
the expected ratio of number of events into $\gamma\gamma$
for SM4 over SM3
is about $1.5-2.5$~at leading order~\cite{Guo:2011ab,Ruan:2011qg}.
However, a global fit to all relevant observables
(Higgs searches and electroweak precision data),
assuming Dirac neutrinos and a Higgs mass of 125 GeV,  shows
that data are better described by the SM3~\cite{Eberhardt:2012sb}. 
On the other hand, as commented above, within the SM4 the cancellations in
the $\gamma\gamma$ channel
at leading order render next-to-leading order radiative corrections important.
These corrections tend to decrease even further the two-photon production rate $\sigma(gg\rightarrow H)\times BR(H\rightarrow \gamma\gamma)|_{SM4}$.
Therefore, were the 125 GeV Higgs hint confirmed, by
combining the $\gamma\gamma$, $ZZ^*$, $WW^*$ and the $f\bar{f}$ channels
a perturbative SM4 \emph{with just one SM Higgs doublet} would be excluded, even in the case
$m_{\nu_{4}}<m_{W}$ \cite{Djouadi:2012ae,Kuflik:2012ai}.
Otherwise, in principle it seems possible that if
 $m_{\nu_{4}}\lesssim m_{W}$ and $\nu_4$ is long-lived,
some portion of the low Higgs
mass parameter space, previously allowed to be between $114-160$ GeV,
is still allowed by the new data.
Moreover, if one does not trust the convergence of perturbation theory
in the $\gamma\gamma$ channel and drops it from the global analysis, including Higgs searches, $R_b$  and oblique parameters, the SM4 
with Dirac neutrinos is strongly constrained but still viable~\cite{Buchkremer:2012yy}. Considering neutrino Majorana masses will presumably open
up even more the allowed parameter space of the model. 

The previous bounds from LEP on the masses of unstable (in collider sense) fourth-generation neutrinos 
were $m_{\nu_{4}}>62.1$ GeV. 
Using  CDF  inclusive like-sign dilepton analysis, $\nu_4$ masses below
$m_W$ can be excluded for Higgs masses up to $2 m_W$ \cite{Carpenter:2011cm},
therefore in this case  the ATLAS and CMS analysis for the Higgs boson still apply, and
at least  the range $120-600$ GeV for a SM4 Higgs boson is excluded.

To know definitely  whether the SM4 Higgs boson is excluded or not, we will have
to wait for new data and a combined analysis of the different channels, 
$\gamma\gamma$, $ZZ^*$, $WW^*$ and $f\bar{f}$,  including correctly all radiative
corrections. However, even if the SM-like four-generation Higgs is excluded, many possibilities may arise
in extensions of a four-generation scenario, for instance, with an extra Higgs doublet (see
\cite{Chen:2012wz,Bellantoni:2012ag} where the observed signatures of LHC are explained in the framework of 4G two-Higgs-doublet models).

\section{Conclusions\label{conclusions}}

We have addressed the question of the generation and nature of neutrino masses in the 
context of the SM with four families of quarks and leptons.

The three light neutrinos can obtain their masses from a variety of mechanisms with or without
new neutral fermions, but the huge hierarchy among such masses and those of the remaining 
fermions is more naturally explained assuming that they have Majorana nature.

On the other hand, current bounds on fourth-generation neutrino masses  imply that, 
although in principle the same mechanisms are also available, 
most of them are not natural or provide too small fourth-generation
neutrino masses; therefore,  we have argued that at least one right-handed neutrino is needed.
This would suggest 
that, contrary to the light neutrinos, fourth-generation ones are naturally Dirac.  

However,  we have shown that  if lepton number is not conserved in the light neutrino sector,
the right-handed neutrino must have a Majorana
mass term whose size depends on the underlying mechanism for LNV, unless 
Yukawa couplings of the light leptons to the right-handed neutrino are forbidden. 
We have estimated the natural size of such Majorana mass term within two frameworks 
for the light neutrino masses, namely see-saw type I and type II.
We have seen that,    
even if we set it to zero by hand in the Lagrangian at tree level, it is
generated at two-loops , and
although it depends on the Yukawa couplings and the LNV scale responsible for
light neutrino masses, it can be up to the TeV scale. We have developed a model
where this Majorana mass is forbidden at tree level by a global symmetry, and it
is generated radiatively and finite once this symmetry is broken spontaneously
(see appendix \ref{app-A}).

We have then considered a minimal four-generation scenario, with neutrino Majorana
masses parametrized by the Weinberg operator and one right-handed 
neutrino $\nu_{\mathrm{R}}$, which has Yukawa couplings to the four lepton doublets 
and non-zero Majorana mass. 
We have analyzed the phenomenological constraints  on the parameter space of such 
a model,
derived from direct searches for four-generation leptons, universality tests, 
charged lepton flavour-violating processes and neutrinoless double beta decay. 
We have pointed out that the Majorana mass for the fourth-generation 
neutrino induces relatively large
two-loop contributions to the light neutrino masses, which can
easily exceed the atmospheric scale and the cosmological bounds. Indeed, 
this sets the
strongest limits on the masses and mixings  of fourth-generation neutrinos, 
collected in figure~\ref{fig:cosmo}.

To summarize, in the context of a SM with four generations, we have shown that if light neutrinos
are Majorana particles, it is natural that also the fourth-generation neutrino
has the Majorana character. We did so by calculating the fourth-neutrino Majorana masses
induced by the three light neutrino ones.  This has important implications for the neutrino 
and Higgs sectors of these models, which are being actively tested at the LHC.

\section{Acknowledgments}

We thank Enrique Fern\'andez-Mart\'\i nez and Jacobo L\'opez-Pav\'on for 
fruitful discussions on the matter of neutrinoless double beta decay.
This work has been partially supported by the Spanish MICINN under
grants FPA-2007-60323, FPA-2008-03373, FPA2011-23897, FPA2011-29678-C02-01, 
Consolider-Ingenio PAU (CSD2007-00060)
and CPAN (CSD2007- 00042) and by Generalitat Valenciana grants PROMETEO/2009/116
and PROMETEO/2009/128.  A.A.
and J.H.-G. are supported by the MICINN under the FPU program.

\appendix

\section{A model for calculable right-handed neutrino masses} \label{app-A}

In this appendix we present a model which gives a realistic pattern of neutrino masses 
in the context of the SM with four-generations and in which 
the right-handed neutrino mass of the fourth generation
is generated radiatively and finite. This is an illustration of
the general (model-independent) mechanism
discussed in section \ref{SM4} which allowed us to estimate the
size of Majorana neutrino masses for the fourth-generation right-handed
neutrinos if the three light active neutrinos are Majorana particles.

Let us consider the SM with four generations and four right-handed neutrinos 
$\nu_{\mathrm{R} i}$ ($i=1,\cdots,4$). To implement the ordinary see-saw, 
we need three of them very heavy while one of them should be much lighter 
in order to avoid a too light fourth-generation active neutrino. Then, it is 
natural to require that one of the fourth right-handed neutrino is massless
at tree level and let its mass be generated by radiative corrections. For that 
purpose we add three extra chiral singlets $s_{\mathrm{L} a}$
$(a=1,\cdots,3$). In order to break lepton number we will also include a complex 
scalar singlet $\sigma$

We assign lepton number in the following way
\begin{equation}
	\ell_{j} \rightarrow e^{i\alpha} \ell_{j} \;, \quad e_{\mathrm{R} j} \rightarrow e^{i\alpha} e_{\mathrm{R} j} \;, 
		\quad
	\nu_{\mathrm{R} j} \rightarrow e^{i \alpha} \nu_{\mathrm{R} j} \;,
		\quad
	\sigma \rightarrow e^{i \alpha} \sigma \; ;
\end{equation}
the $s_{\mathrm{L} a}$ do not carry lepton number. With these assignments and the requirement 
that lepton number is conserved we have the following Yukawa Lagrangian 
\begin{equation}
	\mathcal{L}_{Y} = -\overline{\ell} \, Y_{e} e_{\mathrm{R}} \phi - 
			\overline{\ell} \, Y_{\nu} \nu_{\mathrm{R}} \tilde{\phi} - 
			\sigma \, \overline{\nu_{\mathrm{R}}} \, y^* \, s_{\mathrm{L}} - 
			\frac{1}{2} \, \overline{s_{\mathrm{L}}^{\mathrm{c}}} M^* 
			s_{\mathrm{L}}	+ \mathrm{H.c.} \; ,
\end{equation}
where $Y_{e}$ and $Y_{\nu}$ are the ordinary four-generation Yukawa couplings,
$y_{ia}$, along this appendix, is a general $4 \times 3$ matrix while $M$ is a symmetric
$3\times3$ matrix, which without loss of generality can be taken
diagonal and positive. We choose the scalar potential in such a way that lepton number
is conserved and subsequently spontaneously broken by the VEV of $\sigma$, $v_\sigma=\langle \sigma\rangle$.  Thus, 
the model will contain a singlet Majoron. Alternatively, we could also choose to softly break lepton number in the 
potential to avoid the Majoron without changing the point we would like to illustrate. 
Before spontaneous symmetry breaking only $s_{\mathrm{L} a}$  are massive.
We will take $M$ very large (around GUT scale). After $\sigma$ gets
a VEV (which is somewhat free, but we can take it just a bit below $M$), we will have a mass matrix 
for the combined system
$\nu_{\mathrm{R}}-s_{\mathrm{L}}$ of see-saw type. 
Therefore, if $y \, v_{\sigma} \ll M$
the four right-handed neutrinos will get a $4 \times 4$ Majorana mass matrix
\begin{equation}
	M_{\mathrm{R}}^{(0)} \simeq v_{\sigma}^{2} \, y M^{-1} y^{\mathrm{T}} \; ;
\end{equation}
this is basically the see-saw formula but applied to the right-handed neutrinos
and changing the VEV of the Higgs doublet for that of the singlet $\sigma$. 
This matrix has
rank 3 and, therefore, only three of the right-handed neutrinos will obtain
a tree-level mass. The other neutrino will remain massless at tree
level. However, at two loops, due 
to the mechanism described in section \ref{SM4},
also the fourth right-handed neutrino
will acquire a Majorana mass. We depict the diagram giving rise to 
this mass in figure \ref{diag-model-complete}; the diagram is obviously finite by power counting
and the generated mass matrix can be estimated as
\begin{equation}
M_{\mathrm{R}}^{(2)} \sim \frac{v_{\sigma}^{2}}{(4\pi)^{4}} \, 
					(Y^\dagger_{\nu} Y_{\nu})^{\mathrm{T}}\, y M^{-1} y^{\mathrm{T}} \,
					Y^\dagger_{\nu} Y_{\nu} \ln\left(\frac{M}{y v_\sigma}\right)\; .
\end{equation}

\FIGURE{
	\centering
	\includegraphics[width=0.6\textwidth]{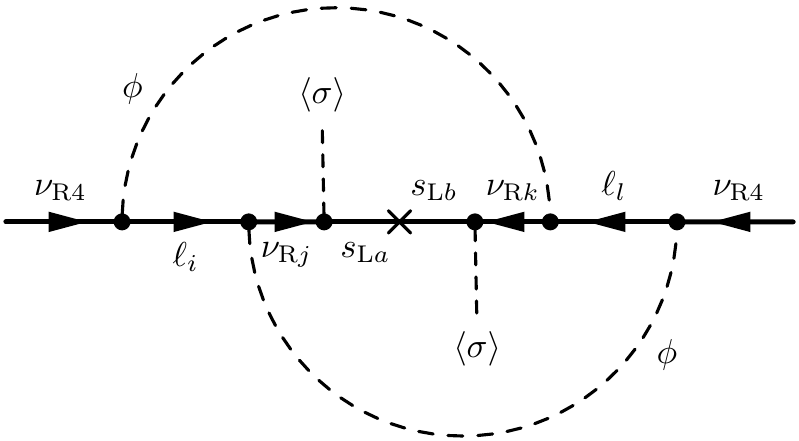}
	\caption{The process which generates masses for the right-handed
			neutrinos at two loops.} \label{diag-model-complete}
}
Since $Y_{e}$ does not enter in these calculations we can choose a
basis in which $Y_{e}$ is arbitrary but $Y_{\nu}$ is diagonal and
real. If we take the logarithm order $1$, $\ln\left(\frac{M}{y v_\sigma}\right)\sim 1$, we see that
$M_{\mathrm{R}}^{(2)}$ is also projective but in a different direction,
given by 
$y^{\prime}= (Y^\dagger_{\nu} Y_{\nu})^{\mathrm{T}} y$;
then we can write
the full right-handed neutrino mass matrix as
\begin{equation}
	M_{\mathrm{R}} \sim v_{\sigma}^{2} \, \left( y M^{-1} y^{\mathrm{T}} 
			+ \frac{1}{(4\pi)^{4}} \, y^{\prime} M^{-1} 
			{y^{\prime}}^{\mathrm{T}} \right) \: ,
\end{equation}
which, in general, has rank 4 and gives a Majorana mass to the fourth right-handed
neutrino. To see how it works, let us discuss a simplified example,
with the following structure for the $s_{\mathrm{L}}$ Yukawas:
\begin{equation}
	y = \begin{pmatrix}
			y_{1} & 0 & 0 \\
			0 & y_{2} & 0 \\
			0 & 0 & y_{3} \\
			0 & 0 & y_{4}
		\end{pmatrix} \: .
\end{equation}
Let us also choose $M$ diagonal and with elements $M_i$; then, at tree level we 
obtain an almost diagonal mass matrix,
\begin{equation}
	M_{\mathrm{R}}^{(0)} = v_\sigma^2 \, \begin{pmatrix}
			\frac{y_{1}^{2}}{M_{1}} & 0 & 0 & 0 \\
			0 & \frac{y_{2}^{2}}{M_{2}} & 0 & 0 \\
			0 & 0 & \frac{y_{3}^{2}}{M_{3}} & \frac{y_{3}y_{4}}{M_{3}} \\
			0 & 0 & \frac{y_{3}y_{4}}{M_{3}} & \frac{y_{4}^{2}}{M_{3}}
	\end{pmatrix} \: ,
\end{equation}
which has a zero eigenvalue. 
At two loops we will have
\begin{equation}
	M_{\mathrm{R}}^{(2)} = \frac{v_{\sigma}^{2}}{(4\pi)^{4}} \, \begin{pmatrix}
		\frac{y_{1}^{\prime2}}{M_{1}} & 0 & 0 & 0 \\
		0 & \frac{y_{2}^{\prime2}}{M_{2}} & 0 & 0 \\
		0 & 0 & \frac{y_{3}^{\prime2}}{M_{3}} & 
			\frac{y_{3}^{\prime}y_{4}^{\prime}}{M_{3}} \\
		0 & 0 & \frac{y_{3}^{\prime}y_{4}^{\prime}}{M_{3}} & 
			\frac{y_{4}^{\prime2}}{M_{3}}
	\end{pmatrix} \: ,
\end{equation}
with $y_{i}^{\prime} = y_{i} {(Y_{\nu})_i}^{2}$, and $(Y_{\nu})_i$ the diagonal
elements of $Y_{\nu}$. $M_{\mathrm{R}}^{(2)}$ has also rank 3. However, 
the sum of $M_{\mathrm{R}}^{(0)}$ and $M_{\mathrm{R}}^{(2)}$
has rank 4, and the fourth $\nu_{\mathrm{R}}$ acquires a mass. We can
estimate it by considering $M_{\mathrm{R}}^{(2)}$ a small perturbation to $M_{\mathrm{R}}^{(0)}$ 
and find that 
\begin{equation}
	m_{\mathrm{R} 4} \sim \frac{v_{\sigma}^{2}}{(4\pi)^{4} M_{3}} \, 
			\frac{y_{4}^{2} \, y_{3}^{2}}{y_{3}^{2}+y_{4}^{2}} \, 
			\left( {(Y_{\nu})^2_4} - {(Y_{\nu})^2_3} \right)^{2} \: ,
\end{equation}
while the mass of the third right-handed neutrino is of order (the other two are also order $y^2 v^2_\sigma/M$ as 
can be seen from the mass matrix)
\begin{equation} \label{mR3-mass}
	m_{\mathrm{R} 3} \sim (y_{3}^{2}+y_{4}^{2}) \, \frac{v_{\sigma}^{2}}{M_{3}} \: .
\end{equation}
Therefore, if we rewrite the fourth-generation right-handed neutrino mass $m_{\mathrm{R} 4}$ in terms of
$m_{\mathrm{R} 3}$ we have
\begin{equation} \label{mR4-mass-rough}
	m_{\mathrm{R} 4} \sim \frac{m_{\mathrm{R} 3}}{(4\pi)^{4}} 
			\frac{y_{4}^{2} \, y_{3}^{2}}{(y_{3}^{2}+y_{4}^{2})^2} \, 
			\left( {(Y_{\nu})^2_4} - {(Y_{\nu})^2_3} \right)^{2}  \: ,
\end{equation}
which is roughly the structure that one would expect from the effective
theory obtained by integrating the new fermions $s_{\mathrm{L} a}$, {\em i.e}, $m_{\mathrm{R} 4}$ obtains a 
contribution proportional to the heavy right-handed 
Majorana masses $m_{\mathrm{R} 3}$ suppressed by a two-loop factor and Yukawa couplings. After all, the diagram
in figure~\ref{diag-model-complete} reduces to the diagram in figure~\ref{2loop-seesaw1} when the fermion lines of $s_{\mathrm{L} a}$ 
are contracted to a point. The result also shows that, as expected,  the exact coefficient depends on the details of the model.
These expressions could be generalized to a more general structure of Yukawa couplings, leading to similar,
although more complicated expressions. 

As for other features of this model, we will just mention that 
as lepton number is broken spontaneously, a Majoron will appear. Since the Majoron is a singlet and
$v_{\sigma}$ is large their couplings to standard model particles are suppressed and, therefore, this Majoron
should not create  any problem. On the other hand, it could have some advantages in cosmological contexts;
if lepton number is also broken softly (for instance with a mass
term $\sigma^{2}$) the Majoron will become a massive pseudo-Majoron,
which could constitute a good dark matter candidate.

In any case, this simple example illustrates how the general mechanism discussed in section \ref{SM4} works
in a complete renormalizable model; if $m_{\mathrm{R} 4}$ is zero at tree level and light neutrinos are Majorana 
(therefore lepton number is not conserved), in general $m_{\mathrm{R} 4}$ will be generated at two-loops with
the behaviour discussed in section \ref{SM4}.

\newcommand{\url}[1]{\href{#1}{{\tt #1}}}

\providecommand{\href}[2]{#2}\begingroup\raggedright\endgroup

\end{document}